\documentclass[sigconf, nonacm, 9pt]{acmart}
\setcopyright{none}
\acmYear{2026}
\acmISBN{}
\acmDOI{}
\startPage{1}

\setcopyright{none}

\AtBeginDocument{%
  }

\usepackage{amsmath}
\usepackage{cleveref}

\usepackage{amssymb}
\usepackage{graphicx}
\usepackage{tabularx}
\usepackage{array}
\usepackage{microtype}
\usepackage{amsfonts}
\usepackage{textcomp}
\usepackage{url}
\usepackage{xspace}
\usepackage{adjustbox}
\usepackage{hyperref}
\usepackage{mathtools}
\usepackage{lscape}
\usepackage{afterpage}
\usepackage{relsize}
\usepackage{enumitem}
\usepackage{subcaption}
\usepackage{booktabs}

\newcommand{\acro}[1]{{\textsmaller[0.5]{#1}}}
\ifdefined\shortcite\relax\else\newcommand{\shortcite}[1]{\cite{#1}}\fi
\ifdefined\nd\relax\else\newcommand{\nd}[1]{#1{\acro{D}}}\fi
\newcommand{\graphblas}{Graph\acro{BLAS}\xspace}

\newcommand{\opsfrac}[2]{\left\langle
  \raisebox{0.3ex}{\footnotesize$\frac{#1}{#2}$}
\right\rangle}

\usepackage{caption}
\usepackage{float}
\newfloat{cascade}{hbtp}{lop}
\floatname{cascade}{Cascade}
\usepackage{mdframed}

\newcommand{\eqcomment}[1]{\text{\scriptsize\,[#1]}\notag\\}

\newenvironment{compactalign}
  {\begingroup
   \setlength{\abovedisplayskip}{6pt}%
   \setlength{\belowdisplayskip}{6pt}%
   \setlength{\jot}{2pt}%
   \align}
  {\endalign\endgroup}

\setlist[itemize]{topsep=2pt,partopsep=0pt,parsep=0pt,itemsep=1pt}
\setlist[enumerate]{topsep=2pt,partopsep=0pt,parsep=0pt,itemsep=1pt}

\makeatletter
\let\hyxmp@parse@acmart\relax
\let\hyxmp@prism@schema\relax
\makeatother

\author{Toluwanimi O. Odemuyiwa}
\affiliation{%
  \institution{University of California, Davis}
  \city{Davis}
  \state{California}
  \country{USA}
}
\email{todemuyiwa@ucdavis.edu}

\author{Serban D. Porumbescu}
\affiliation{%
  \institution{University of California, Davis}
  \city{Davis}
  \state{California}
  \country{USA}
}
\email{sdporumbescu@ucdavis.edu}

\author{Muhammad Osama}
\affiliation{%
  \institution{University of California, Davis}
  \city{Davis}
  \state{California}
  \country{USA}
}
\email{mosama@ucdavis.edu}

\author{Joel S. Emer}
\affiliation{%
  \institution{Massachusetts Institute of Technology}
  \city{Cambridge}
  \state{Massachusetts}
  \country{USA}
}
\email{emer@csail.mit.edu}

\author{John D. Owens}
\affiliation{%
  \institution{University of California, Davis}
  \city{Davis}
  \state{California}
  \country{USA}
}
\email{jowens@ucdavis.edu}

\begin{document}

\title{The Einsum-Enabled Design Space for Graph Algorithms: A BFS Case Study}
\renewcommand\footnotetextcopyrightpermission[1]{}
\fancyhead{}
\begin{abstract}

We propose a principled approach to reasoning about various graph algorithm implementations.
We leverage the extended general Einsum notation (EDGE) which allows us to factor complexity along four axes: algebraic manipulation, mapping, format, and low-level implementations.
Using breadth-first search (BFS) as a driving example and case study, we apply our methodology to explore over 90 variations across 26 categories of optimization choices for our GPU-based implementations.
In addition to showing that our approach is general enough to represent previously discovered algorithmic techniques such as the pull variant of BFS, we discover never-before-seen variants that lead to geomean performance benefits ranging from 1.2$\times$ to 1.7$\times$ over the best Gunrock baseline variation for graphs with mid- to high- normalized degree variance.
\end{abstract}

\maketitle

\section{Introduction}
\label{sec:intro}
\begin{figure}
  \centering
  \includegraphics[width=\columnwidth]{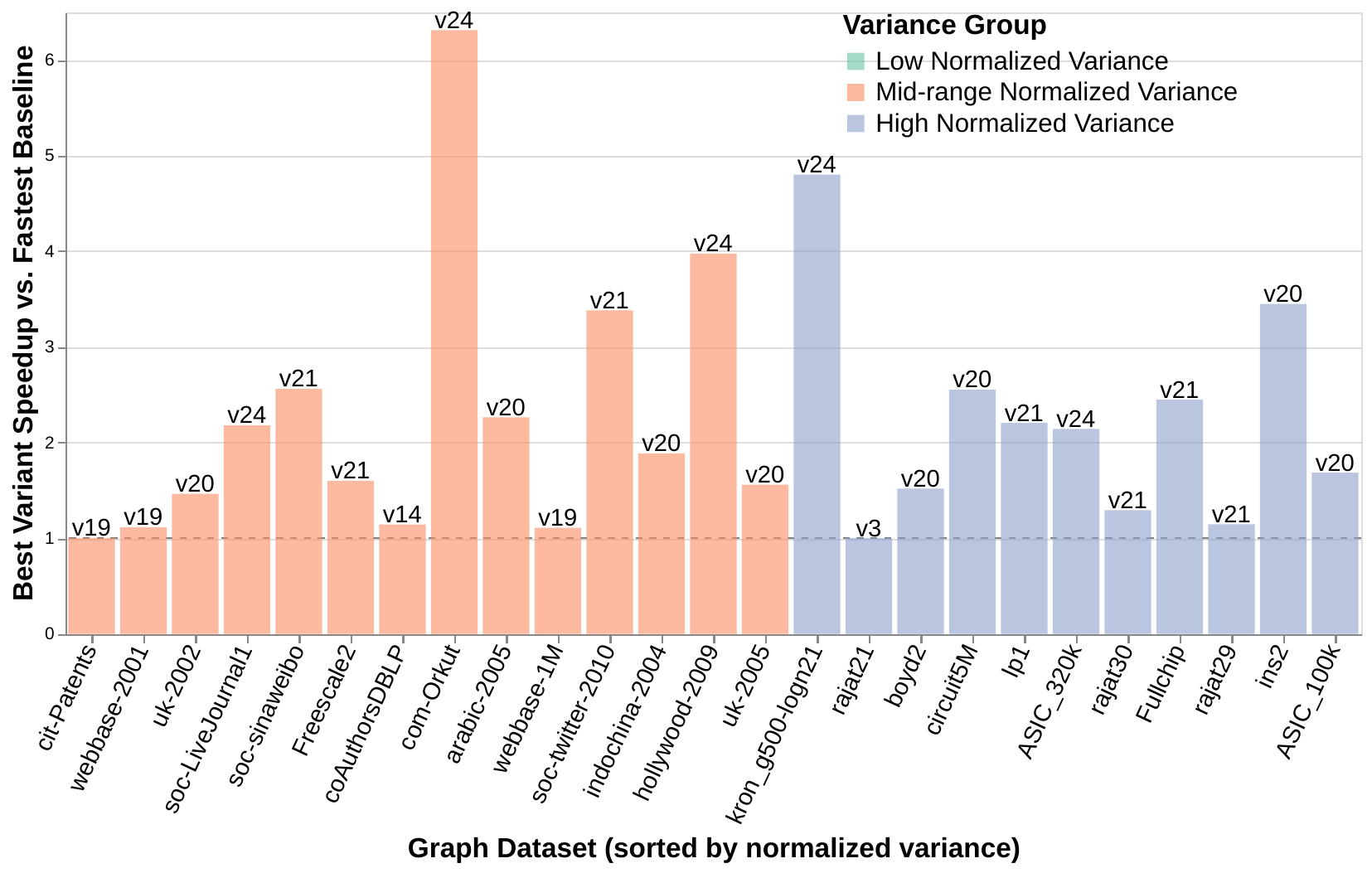}
  \caption{
  The best-case speedup on each graph dataset, normalized to the best-case Gunrock baseline.
  Each label, (e.g., ``v20''), corresponds to a specific BFS implementation, which is generated by applying composable EEDS transformations to a single EDGE specification.
  }
  \label{fig:points-summary}
  \vspace{-0.4em}
\end{figure}

High-performance graph algorithms reside in a rich and combinatorial design space. Implementers must choose among data formats (e.g., COO, CSR), data layouts for auxiliary structures (e.g., queues, bitmasks), computational views
(vertex-centric~\cite{McCune:2015:TLA, Khan:2017:VCG, Emoto:TLV:2016,  Malewicz:2010:PAS},
edge-centric~\cite{Zhou:2018:FFE, Roy:2013:XEC, Zhang:2018:ABG,Zhu:2020:WEC},
matrix-based~\cite{Gilbert:2006:HGA, Yang:2018:IPE, Artiles:2021:TGB, Besta:2017:SSV, Buluc:2017:DGA,Mattson:2019:LAC},
and graph-centric~\cite{Tian:2013:FTL,Fan:2017:FTP}),
partitioning strategies~\cite{Zhang:2018:GAH, Finnerty:2019:DBD,Malewicz:2010:PAS,Salihoglu:2013:GPS,Tian:2013:FTL},
and a variety of mechanisms for assigning work to software and hardware resources~\cite{Merrill:2012:SGG,Osama:2023:PMG,Osama:2022:EOP,Shun:2013:LAL, Zhang:2018:GAH}.
Performance depends heavily on graph characteristics (density, connectedness, size)~\cite{Graph500, Graph500:2017, Malicevic:2017:EYA, Nai:2015:GUG, Wang:2017:GGG, Zhang:2018:GAH} and on how well an implementation leverages optimization opportunities under architectural constraints such as memory hierarchy, parallelism, and communication overhead. As such, optimization is often driven by ad hoc combinations of operational techniques (push versus pull traversal, masking, atomic elimination, load balancing)~\cite{Merrill:2012:SGG, Zhang:2018:GAH, Wang:2017:GGG,Malewicz:2010:PAS,Salihoglu:2013:GPS,Mastrostefano:2013:EBF, McCune:2015:TLA, Green:2021:BBA} and evaluated within specific frameworks.

Although these strategies target the same underlying computation, they are generally treated as algorithmically distinct designs. The dimensions along which implementations differ are entangled and explored empirically. This fragmentation obscures structural relationships between implementations and confounds systematic exploration.

We argue that graph algorithm implementations are better understood as points in a structured transformation space. By expressing algorithms as extended Einsums (\S~\ref{sec:what-is-an-einsum})---specifically, in the recently proposed EDGE language~\cite{Odemuyiwa:2024:edge}---we expose a shared computational core from which diverse implementations arise through composable transformations. In this view, push vs.\ pull traversal corresponds to a set of algebraic and loop transformations, as detailed in \S~\ref{ssec:pushpull}.

To make this structure explicit, we introduce the Einsum-Enabled Design Space (EEDS) in \S~\ref{sec:dse_with_ta}, a structured design space spanned by four orthogonal dimensions: the Extended Einsum Space (algebraic manipulation), Mapping Space, Format Space, and Code Space (low-level implementation). Within EEDS, algebraic structure becomes an explicit dimension along which implementations can be optimized. Our work here uses breadth-first search (BFS) as a case study to illustrate how structured traversal of EEDS exposes and navigates the implementation space of a representative graph primitive. We choose BFS because it has a simple algorithm, a rich history of study, and (as we show) a broad variety of interesting implementation alternatives. The contributions of our work are:

  \begin{enumerate}
    \item We generalize algebraic manipulation as a first-class optimization axis within the space of Einsum-based implementations (\S~\ref{sec:dse_with_ta}). To our knowledge, this is the first work in tensor algebra to generalize this dimension~\cite{Nayak:2024:FML_micro}.

    \item We define the Einsum-Enabled Design Space (EEDS), a structured framework spanning algebraic, mapping, format, and low-level code transformations. We show how widely used graph implementation techniques correspond to specific regions of this space, exposing structural relationships that are typically treated as unrelated designs (\S\S~\ref{sec:dse_with_ta},~\ref{sec:essentials}).

    \item Using GPU breadth-first search (BFS) as a case study, we systematically traverse EEDS by applying composable transformations to a single EDGE specification. This recovers prior BFS variants and uncovers new implementations that achieve up to a $1.7\times$ geometric-mean speedup over the best-case Gunrock baseline on large, high-degree-variance graphs. We evaluate 26 resulting BFS variations and summarize their performance in Figure~\ref{fig:points-summary}. Selecting the best variant for each graph yields a performance improvement of 1.97$\times$ on mid-variance to high-variance graphs. (\S~\ref{sec:performance}).
\end{enumerate}

Overall, we use BFS---one of the fundamental traversal primitives underlying many graph algorithms---as a representative case study.
Our goal is not to exhaustively enumerate the entire implementation space, nor to claim that every newly identified variant will outperform existing baselines.
Rather, we demonstrate that graph implementations can be organized within a structured  design space, making previously implicit optimization dimensions explicit and systematically traversable.
Performance improvements serve as evidence that this structured exploration can uncover meaningful new implementations.

\paragraph{What This Work Is Not.}
This work does not aim to replace graph frameworks such as Gunrock~\cite{Wang:2016:GAH}, IrGL~\cite{Pai:2016:ACF}, or GraphIt~\cite{Zhang:2018:GAH}.
Rather, we position these systems as point implementations within a broader Einsum-based design space.
Given a specification in EDGE together with mapping, format, and low-level implementation choices, implementations can be generated manually or automatically.
Prior work has demonstrated this methodology for tensor algebra workloads~\cite{Nayak:2023:TDF, Nayak:2024:FML_micro}.

\section{Background}
\subsection{Tensor Algebra}
\label{ssec:ODE}
To express graph algorithms, we use a tensor algebra formulation and follow the terminology proposed in TeAAL~\cite{Nayak:2023:TDF} and the ExtenDed General Einsums Language (EDGE)~\cite{Odemuyiwa:2024:edge}.
A tensor is a multidimensional array, where a scalar is 0-D, a vector is 1-D, a matrix is 2-D, and so on.
Einstein summations (Einsums)---or ``tensor index notation''---is the notation of choice for tensor algebra computations~\cite{Kjolstad:2017:TTA}.
For example, the Einsum for matrix multiplication between two input 2-D tensors, $A$ and $B$, is:
\begin{align}
  \label{eqn:einsum:ode}
  Z_{m,n} &= A_{m,k} \times B_{n, k}
\end{align}
According to the operational definition of an Einsum~\cite{Nayak:2023:TDF,Odemuyiwa:2024:edge}, this Einsum specifies the tensors $A$, $B$, and $Z$ and their corresponding dataspaces (associated data values).
Each tensor has \emph{shapes} (maximum sizes) of $M \times K$, $N \times K$, $M \times N$, respectively.
A \emph{rank} refers to the axes or indices of a given tensor~\cite{Nayak:2023:TDF} (e.g., $A$ has ranks $M$ and $K$).
The Einsum also indicates the iteration space. This is composed of the Cartesian product of all the legal coordinates (subscript variables) in the expression.
The iteration space is $[0, M) \times [0, N) \times [0, K)$.
We indicate a $point$ in that iteration space in lower case letters, e.g.: $(m,n,k)$.
An Einsum implementation traverses each $(m, n, k)$ point in the iteration space, performing the operation ($\times$) for the point specified on the right-hand side of the expression.
In this case, it multiplies a scalar from $A$ at location $m,k$ with a scalar from $B$ at location $n,k$.
It then places the result in the output tensor, $Z$, at location $m,n$.
If a result already exists at that location, it reduces (with $+$) the new result into the existing value.

Note that this Einsum does not specify \emph{how} to traverse the iteration space, only which computations to perform.
Instead, we specify the ``how'' separately as a \emph{schedule}~\cite{Kjolstad:2017:TTA} or \emph{mapping}~\cite{Parashar:2019:TSA}.

\subsection{Extended Einsums for Graph Workloads}
\label{sec:what-is-an-einsum}
TeAAL enables the expression of multi-phase workloads using a \emph{cascade} of Einsums: a DAG of interdependent Einsum expressions.
EDGE further extends Einsums to support general workloads, including graph algorithms~\cite{Odemuyiwa:2024:edge}.
We now summarize the key extensions we use in this work:
\subsubsection{Iterative Computations.}
EDGE introduces \emph{iterative (or generational) ranks} to express recursive/iterative algorithms.
For example, we can express inclusive scan iteratively~\cite{Nayak:2024:FML_micro}:

\begingroup
\setlength{\abovedisplayskip}{6pt}
\setlength{\belowdisplayskip}{6pt}
\setlength{\jot}{2pt}

\begin{compactalign}
Z_{0} &= 0 \label{eq:einsum:scan:init} \\
Z_{i} &= A_{i-1} + Z_{i-1} \label{eq:einsum:scan:main} \\
&\diamond : i > K \label{eq:einsum:scan:stop}
\end{compactalign}

\endgroup

Rank variable $i$ is iterative, incrementing by one on each pass of the Einsum.
The $\diamond$ symbol indicates the loop's stopping condition, which in this case is $i > K$.

\subsubsection{User-Defined Functions}
EDGE allows user-defined functions (UDFs) to be used as operators in Einsums.
To do this, EDGE defines two ``actions'': \emph{Map} and \emph{Reduce}.
A \emph{Map} applies a UDF to each point in the iteration space (e.g., $\times$ in Equation~\ref{eqn:einsum:ode}),
while a \emph{Reduce} combines the current output value (for a given iteration space point) with the currently computed result at that point ($+$ in Equation~\ref{eqn:einsum:ode}).

To expose optimization opportunities for sparsity, each action consists of two operators:
\emph{Merge} decides which iteration space points to cull, while \emph{Compute} is any user-defined function that performs a computation on the values at the selected iteration space points.
In this work, we use two merge operators: intersect ($\cap$) and union ($\cup$).
Intersect only keeps iteration space points that exist in both input tensors, while union keeps iteration space points that exist in any input tensor.
Table~\ref{tab:compute} lists the common compute operators we use in this work.

While the EDGE work uses a post-fix notation to express Map and Reduce, for compactness we instead use an infix notation with Map on top ($\times(\cap)$) and Reduce on the bottom ($+(\cup)$).
Thus, we can write sparse-sparse GEMM in EDGE as:
\begin{equation}
  \label{eqn:einsum:sparse_gemm}
  Z_{m,n} = A_{m,k} \opsfrac{\times(\cap)}{+(\cup)}  B_{n, k}
\end{equation}
That is, only multiply ($\times$) matching values in $A$ and $B$  when their corresponding coordinates survive intersection ($\cap$, $k$ exists in both $A$ and $B$ for a given $(m, n)$).
We then sum ($+$) the multiplication results into $Z$, if at least one of the current $Z$ and the incoming computed value is non-zero ($\cup$).

\begin{table}
  \caption{Common compute operators used in this paper.}
  \label{tab:compute}
  \centering
  \begingroup
  \small
  \setlength{\tabcolsep}{4pt}
  \renewcommand{\arraystretch}{0.85}

  \begin{tabular}{@{}l p{0.65\columnwidth}@{}}
    \toprule
    \textbf{Operator (Symbol)} & \textbf{Meaning} \\ \midrule
    $\text{AND}$ & Boolean AND operator \\
    $\text{OR}$ & Boolean OR operator \\
    $\text{ANY}$ & Select ANY of the input values \\
    $>$ & Boolean “greater than” operator \\
    $\leq$ & Boolean “less than or equal to” operator \\
    $\rightarrow$ & Select the value in the right-hand operand \\
    $\leftarrow$ & Select the value in the left-hand operand \\
    $\min$ & Select the minimum value across all inputs \\
    $+$ & Add the input values \\
    $\lll$ & Update: take right operand if non-empty, else left operand \\
    \bottomrule
  \end{tabular}

  \endgroup
  \vspace{-0.4em}
\end{table}

\section{Related Work}\label{sec:why_edge}

Developers face many design choices when implementing graph algorithms on a given system.
An abstraction suitable for systematic exploration of these choices should satisfy three properties:
(1) separation of concerns between computation and implementation,
(2) platform independence at the abstraction level, and
(3) sufficient expressiveness to capture diverse graph execution strategies within a unified formulation.
Prior work has advanced subsets of these properties, but no existing approach satisfies all three simultaneously.

GraphIt~\cite{Zhang:2018:GAH} represents one approach to structuring the graph implementation space. It separates the \emph{algorithm}, which specifies the high-level computation, from the \emph{schedule}, which specifies composable optimizations such as traversal order, parallelization strategy, frontier representation, and data layout.
This separation improves clarity and enables controlled exploration of certain implementation choices. However, GraphIt's abstractions remain platform-dependent. Targeting GPUs required Brahmakshatriya et al.\ to introduce G2~\cite{Brahmakshatriya:2021:CGA}, which adds GPU-specific scheduling primitives. As a result, mapping concerns are not fully decoupled from hardware details, limiting portability across architectures.

Linear-algebraic formulations such as \graphblas~\cite{Kepner:2011:GAL, Szarnyas:2021:LLA} leverage the well-known duality between graph traversal and sparse matrix operations.
For example, traversing outbound edges corresponds to multiplication (under a Boolean semiring) of the adjacency matrix by a vertex vector, while selecting unvisited vertices corresponds to masking or filtering the result.
This algebraic view provides mathematical generality and unifies many graph algorithms under a common formalism.
However, in \graphblas, computation and execution strategy remain intertwined. Traversal order, compute granularity, and data access patterns are largely determined by the matrix abstraction itself, limiting explicit reasoning about alternative mappings.
In contrast, Einsum formulations expose a broader implementation space as structured transformations that can be explored independently of a fixed matrix interface.

In the tensor-accelerator domain, TeAAL~\cite{Nayak:2023:TDF} separates tensor computation into four concerns: problem specification, mapping, format, and architecture binding. This separation has enabled systematic exploration of hardware design spaces for tensor algebra workloads. However, TeAAL targets accelerator design and does not directly address graph algorithm implementations in software.

Building on TeAAL and EDGE~\cite{Odemuyiwa:2024:edge}, we extend this separation-of-concerns philosophy to graph algorithms in software.
We identify four analogous concerns (see Figure~\ref{fig:arch_constraints}): (A) Einsum Space (\S\ref{ssec:algebraic_manipulations}), (B) Mapping Space (\S\ref{ssec:space_time}), (C) Format Space (\S\ref{ssec:format}), and (D) Code Space (\S\ref{ssec:codespace}).
We call this structured design space the \emph{Einsum-Enabled Design Space} (EEDS). Each concern defines a set of design knobs which we can adjust independently, enabling principled exploration of implementation choices.

Under this abstraction, platform-specific optimizations can be cleanly separated from platform-independent transformations. For example, GPU-dependent load-balancing strategies in G2 correspond to partitioning and flattening transformations within the Mapping Space of EEDS. Likewise, GraphIt's scheduling choices can be expressed as specific configurations of EEDS knobs\footnote{See Table~\ref{table:graphit_edge} for a mapping of GraphIt options to EEDS concerns.}. In this sense, EEDS captures prior implementation strategies as points within a unified and structured design space.

\begin{table}[t]
  \caption{Example GraphIt optimization alternatives~\cite[Table 1]{Zhang:2018:GAH} and the corresponding design points in EDGE (labels from \S~\ref{sec:dse_with_ta}).}
  \label{table:graphit_edge}
  \centering
  \scriptsize
  \resizebox{\columnwidth}{!}{%
  \begin{tabular}{ll}
    \toprule
    \textbf{GraphIt Option} & \textbf{EDGE Design Point} \\
    \midrule \midrule
    DensePull &
      \begin{tabular}[c]{@{}l@{}}B.1: $d \rightarrow s$\\ C@: leader-follower intersection, $G$ leader\end{tabular} \\ \midrule
    DensePush &
      \begin{tabular}[c]{@{}l@{}}B.1: $s \rightarrow d$\\ D@: leader-follower intersection, $G$ leader\end{tabular} \\ \midrule
    SparsePush &
      \begin{tabular}[c]{@{}l@{}}B.1: $s \rightarrow d$\\ C@: leader-follower intersection, $F$ leader\end{tabular} \\ \midrule
    SparsePull &
      \begin{tabular}[c]{@{}l@{}}B.1: $d \rightarrow s$\\ D@: leader-follower intersection, $P$ leader\end{tabular} \\ \midrule
    DensePull-SparsePush & Instantiate both mapping points \\ \midrule
    bitvector &
      C@: data format choice (see~\cite{Chou:2018:FAF}) \\ \midrule
    vertex-parallel & B.5: parallel on $s$ or $d$ \\ \midrule
    edge-aware-vertex-parallel &
      \begin{tabular}[c]{@{}l@{}}A.6: flatten $s,d \rightarrow (s,d)$\\ A.5: partition $(s,d)$ rank\\ B.5: parallel on each partition\end{tabular} \\ \midrule
    edge-parallel &
      \begin{tabular}[c]{@{}l@{}}A.6: flatten $s,d \rightarrow (s,d)$\\ B.5: parallel on flattened rank\end{tabular} \\ \midrule
    cache partitioning &
      \begin{tabular}[c]{@{}l@{}}A.6: partition on $s$ or $d$\\ B.5: sequential per partition\end{tabular} \\ \midrule
    NUMA partitioning &
      \begin{tabular}[c]{@{}l@{}}A.6: partition on $s$ or $d$\\ B.5: parallel per partition\end{tabular} \\ \midrule
    kernel fusion & B.2, B.3: loop fusion and fission \\
    \bottomrule
  \end{tabular}%
  }
  \vspace{-0.4em}
\end{table}

We therefore view EDGE not as a competing framework, but as a unifying abstraction: one that preserves algebraic generality, separates algorithm from execution strategy, and remains portable across platforms by confining hardware-specific optimizations to the lowest level of binding.

\section{Expressing BFS with EDGE}
\label{sec:bfs_einsums}
\label{ssec:expressing-bfs-with-edge}

\begin{figure}[t]
\centering
\includegraphics[width=\columnwidth]{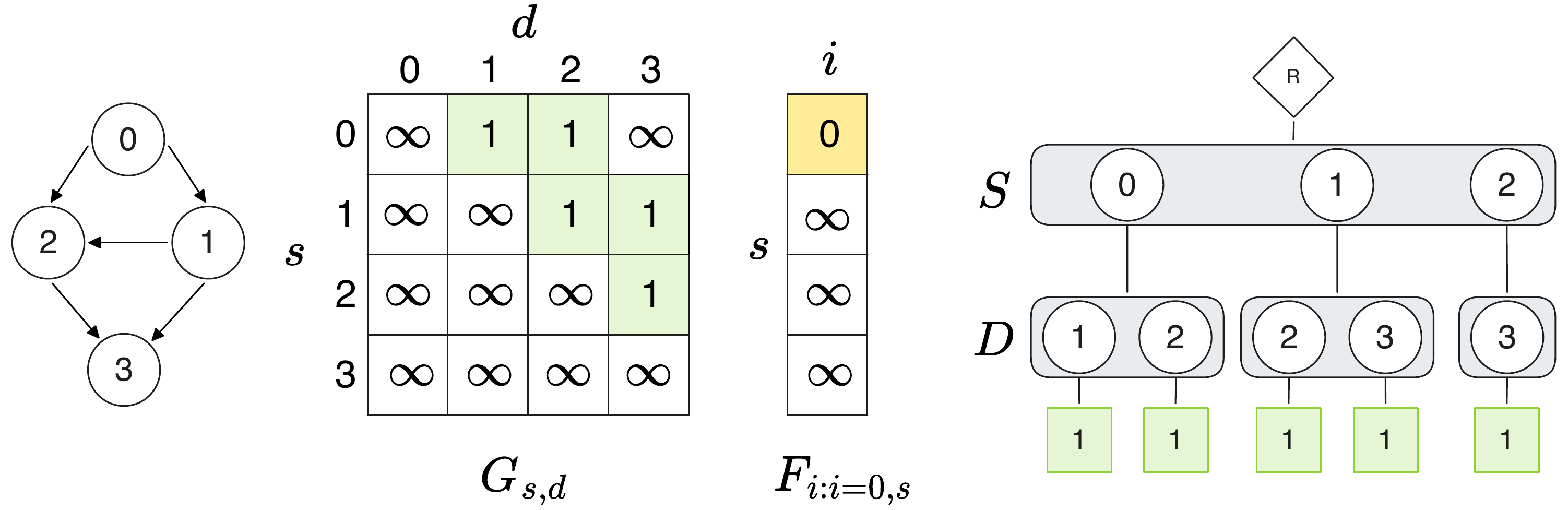}
\caption{Fibertree form of graph G with named ranks S and D.}
\label{fig:mini_graph_fibertree}
  \vspace{-0.4em}
\end{figure}

BFS computes a graph traversal on an unweighted graph.
It is the foundation on which other problems like shortest-path algorithms, centrality measures, and PageRank, amongst others, are built~\cite{Merrill:2012:SGG, Graph500, Che:2009:RAB, Nai:2015:GUG}.
Suppose we have an input graph $G = (V, E)$, where $V$ is the set of all vertices in the graph, $|V|$ is the number of vertices in the graph, and $E$ is the set of all edges.

We describe a directed edge, $e \in E$, by a tuple of vertices $(s \in V, d \in V)$, with the edge starting at $s$ and ending at $d$.
An edge in an undirected graph consists of two directed edges: $(s,d)$ and $(d,s)$.
Given a starting query vertex, graph traversal seeks to find all vertices in $G$ reachable from that query, recording the minimum number of hops, or \emph{depth}, required to reach each vertex.
BFS performs this traversal in a breadth-first manner on an unweighted graph, where vertices at depth $i$ are discovered before vertices at depth $i+1$.
The vertices discovered at depth $i$ form the \emph{active vertex set} or \emph{frontier}~\cite{Shun:2013:LAL,Wang:2017:GGG}.

A successful Einsum must express multiple different aspects of the BFS computation: (1a) find vertices reachable from the current active set and (1b) update their depths;
(2) filter out vertices that have already been discovered; and (3) update the set of visited vertices.
These three steps repeat several times until the active set is empty---that is, all reachable vertices have been discovered.
Cascade~\ref{eqn:edge_bfs} shows the corresponding EDGE BFS expression.

\begin{cascade}[t]
  \caption{EDGE BFS cascade}
  \label{eqn:edge_bfs}
  \centering
  \begin{minipage}{\columnwidth}
  \begin{mdframed}[linewidth=0.6pt,roundcorner=2pt,
                  innertopmargin=4pt,innerbottommargin=4pt,
                  innerleftmargin=6pt,innerrightmargin=6pt]

  \begingroup
  \setlength{\abovedisplayskip}{6pt}
  \setlength{\belowdisplayskip}{6pt}
  \setlength{\jot}{2pt}

  \begin{subequations}
    \begin{align}
      &\triangleright\text{Tensor Declarations} \notag\\
      G^{S\equiv|V|, D\equiv|V|} &\rightarrow  \text{Int, empty=}0 \label{seqn:G}\\
      F^{I, S\equiv|V|} &\rightarrow \text{Int, empty=}\infty \label{seqn:F}\\
      P^{I, D\equiv|V|} &\rightarrow \text{Bool, empty=False} \label{seqn:P}\\
      &\triangleright\text{Initialization} \notag\\
      &\eqcomment{Initialize the source node(s) to a depth of 0}
      F_{0, s} &= 0, P_{0, s} = \text{True}\\
      &\triangleright\text{Extended Einsum} \notag\\
      F_{i+1, d} &= \Biggr(G_{s, d} \opsfrac{+(\cap)_s}{{\text{ANY}(\cup)}_{s}} F_{i, s}\Biggl)_d
                   \opsfrac{\leftarrow\!(\cap)_d}{^.} \neg P_{i, d} \label{seqn:advance}\\
      P_{i+1, d} &= P_{i, d} \opsfrac{\text{OR}(\cup)_d}{^.} F_{i+1, d} \label{seqn:update}\\
      &\diamond: ||F_{i+1}|| \equiv 0
    \end{align}
  \end{subequations}

  \endgroup

  \end{mdframed}
  \end{minipage}
\end{cascade}

The expression specifies an iteration space of $I \times S \times D$ and is a cascade of two Einsums (\eqref{seqn:advance} and~\eqref{seqn:update}).
The first three expressions specify the shape, data type, and empty value of the tensors (Equations~\eqref{seqn:G}--~\eqref{seqn:P}).
$G_{s,d}$ is a 2D tensor representing the input graph, with a shape of $S \times D$, where $S \equiv D \equiv |V|$; Figure~\ref{fig:mini_graph_fibertree} shows its fibertree form with the named $S$ and $D$ ranks.
Each possible $s$ index refers to all the possible source vertices and each possible $d$ index refers to all the possible destination vertices.
We assume $G_{s,d}$ contains a weight value of $1$ whenever an edge exists between vertices $s$ and $d$, and $0$ otherwise.
Tensor $F_{i, s}$ represents the active set of source vertices for the $i$'th iteration (i.e., at depth $i$).
For a given $i, s$ point in the tensor, the corresponding \emph{value} is either empty if $s$ is not in the active set, or contains its integer depth count otherwise.
$P_{i, d}$ is a boolean tensor that indicates if a vertex $d$ has been visited by the start of iteration $i$.
The outputs, $F_{i+1, d}$ and $P_{i+1, d}$ denote the resulting set of discovered vertices at depth $i+1$ and the total set of visited vertices after $i$ iterations, respectively.

Equation~\eqref{seqn:advance} gathers the neighbor list of the frontier $(F)$, adds the current depth of the sources in the frontier to the edge weight of the neighbors, then removes any neighbors that have already been visited.
Specifically, gather occurs during the map action between $G$ and $F$, where we intersect the $s$ rank ($\cap$).
Once gathered, we apply the compute $+$ operator to the current depth of active vertices (values in $F$) and the resulting $(s,d)$ points in $G$.
Since neighbors ($d$ vertices) may belong to multiple sources, we need to select which source updates a neighbor.
We do this using the reduce action, which merges the neighbor lists into a new frontier by reducing over the $s$ rank.
In this particular expression, the $\text{ANY}$ operator selects any one of the duplicate sources.
We intersect the resulting temporary tensor---produced by the computation within the parentheses (`$\biggr($' and `$\biggl)$')---with the complement of the $P$ tensor ($\neg P$).
Since $P$ is a boolean tensor, and the output needs to contain the new active vertex set with their depths, the compute operator is a ``take-left'' operation ($\leftarrow$), which selects the value in the first operand.
This corresponds to filtering out neighbors that have already been visited.
The second expression in Equation~\eqref{seqn:update} updates the visited tensor with the results of the new active set of vertices, by merging with an intersection, and computing with an $\text{OR}$.

\section{The Einsum-Enabled Design Space}
\label{sec:dse_with_ta}

\begin{figure*}
\centering
\includegraphics[width=.8\textwidth]{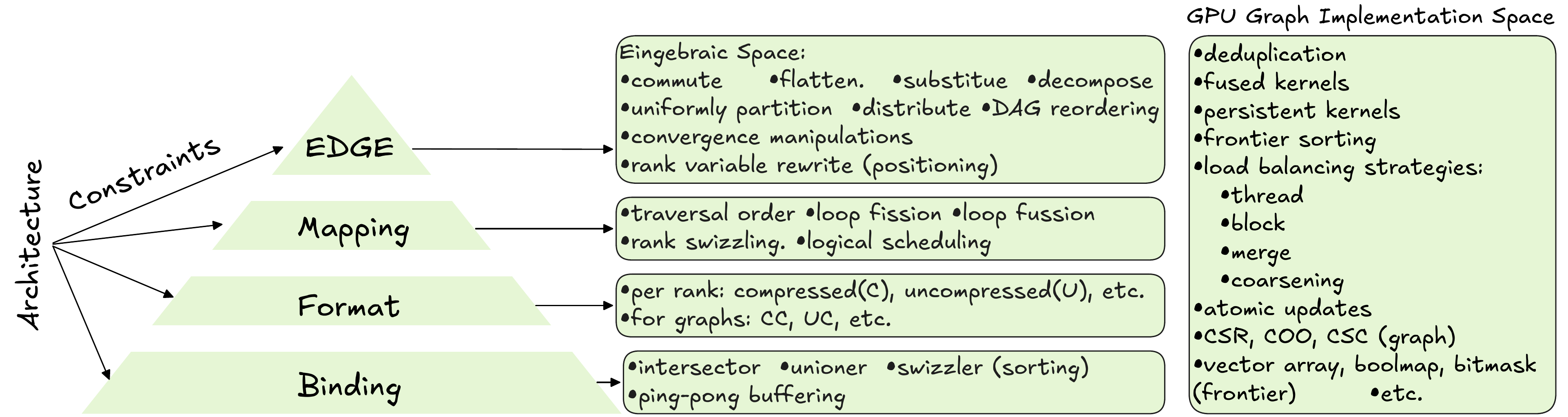}
\caption{The EEDS identifies specifications in the TeAAL pyramid of abstractions. Combined, these correspond to known optimizations / choices in the \@GPU world.}
\label{fig:arch_constraints}
  \vspace{-0.4em}
\end{figure*}

\begin{figure}
  \centering
  \includegraphics[width=0.4\columnwidth]{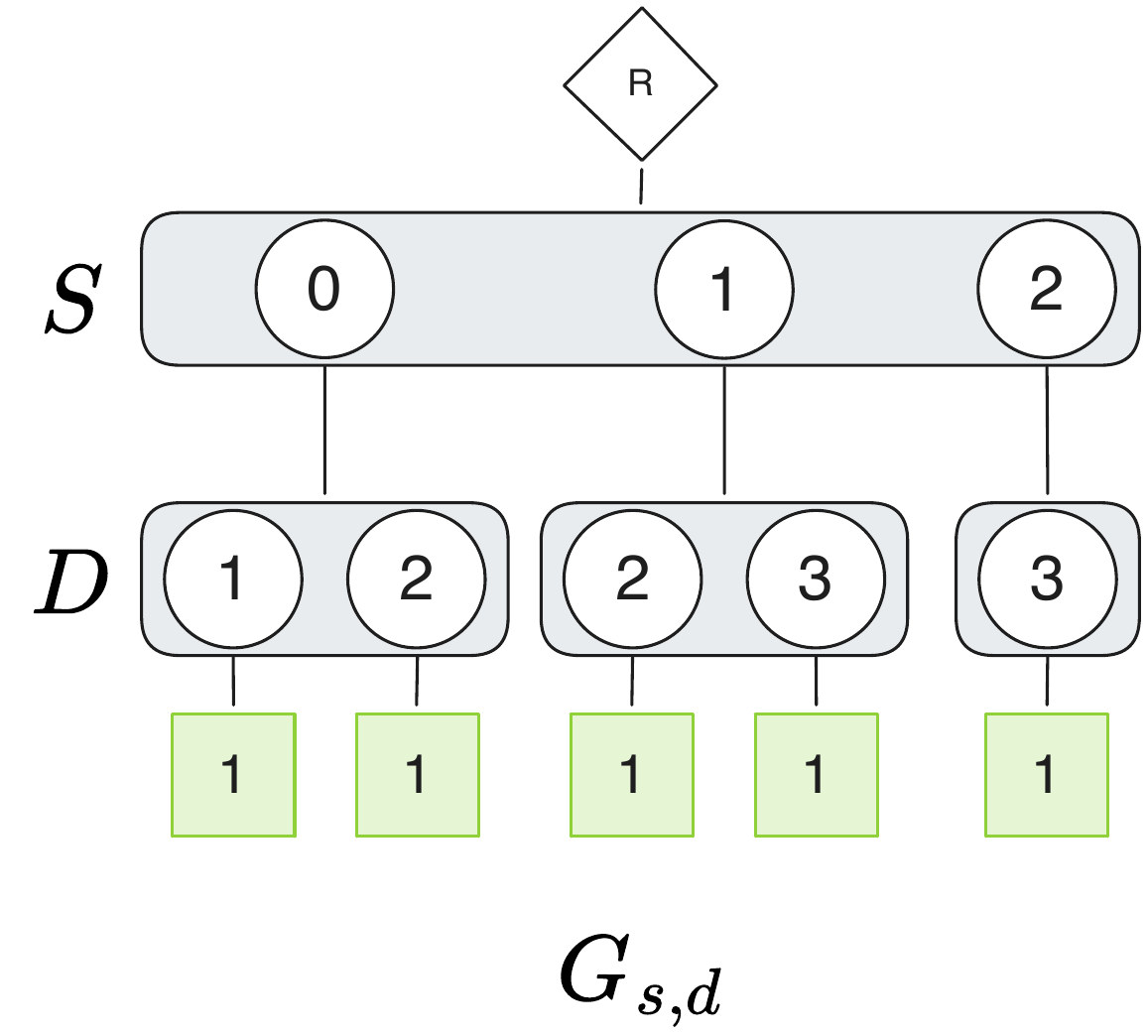}
  \caption{A graph with five edges. Flattening and position shifting on this graph are shown in Figures~\ref{fig:position-middle} and~\ref{fig:position-right}, respectively.}
  \label{fig:position-left}
\end{figure}

\begin{figure}
  \centering
  \begin{subfigure}[t]{0.48\linewidth}
    \centering
    \includegraphics[width=\linewidth, trim=0 0 0 0, clip]
      {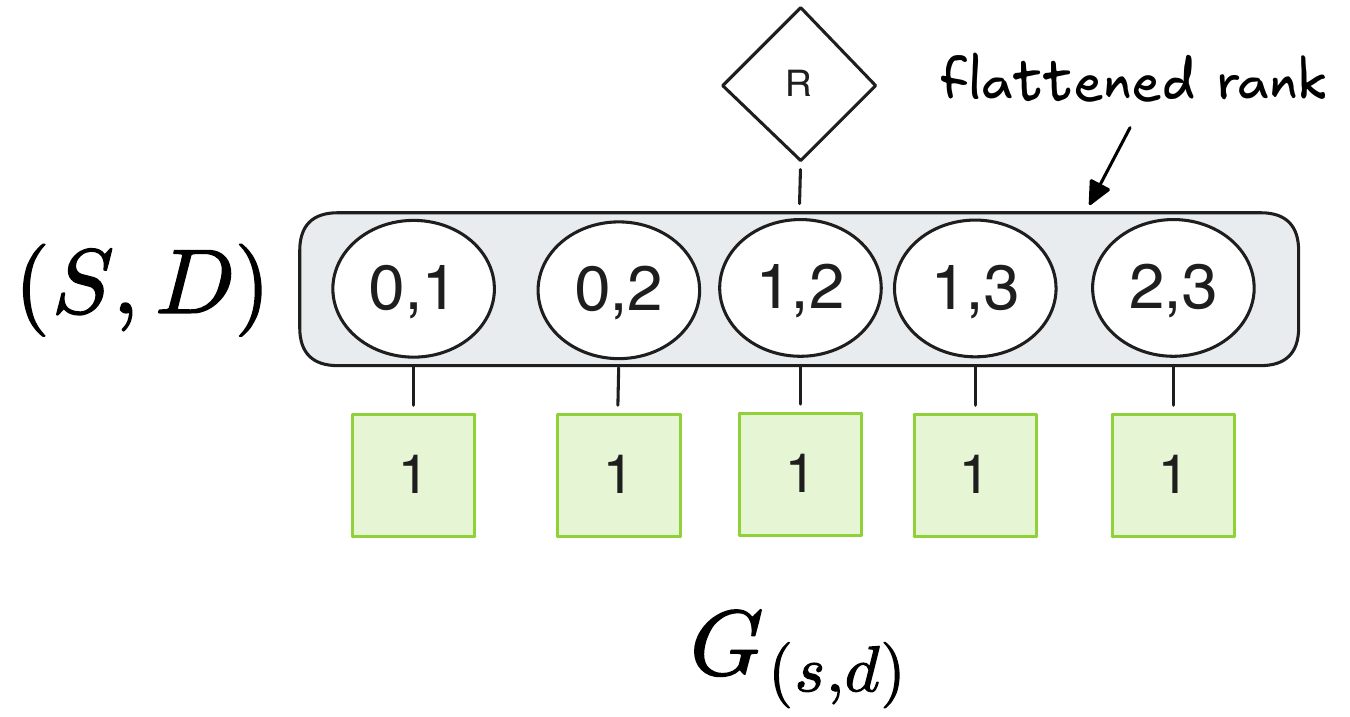}
    \caption{Flattened rank.}
    \label{fig:position-middle}
  \end{subfigure}
  \hfill
  \begin{subfigure}[t]{0.48\linewidth}
    \centering
    \includegraphics[width=\linewidth, trim=0 0 0 0, clip]
      {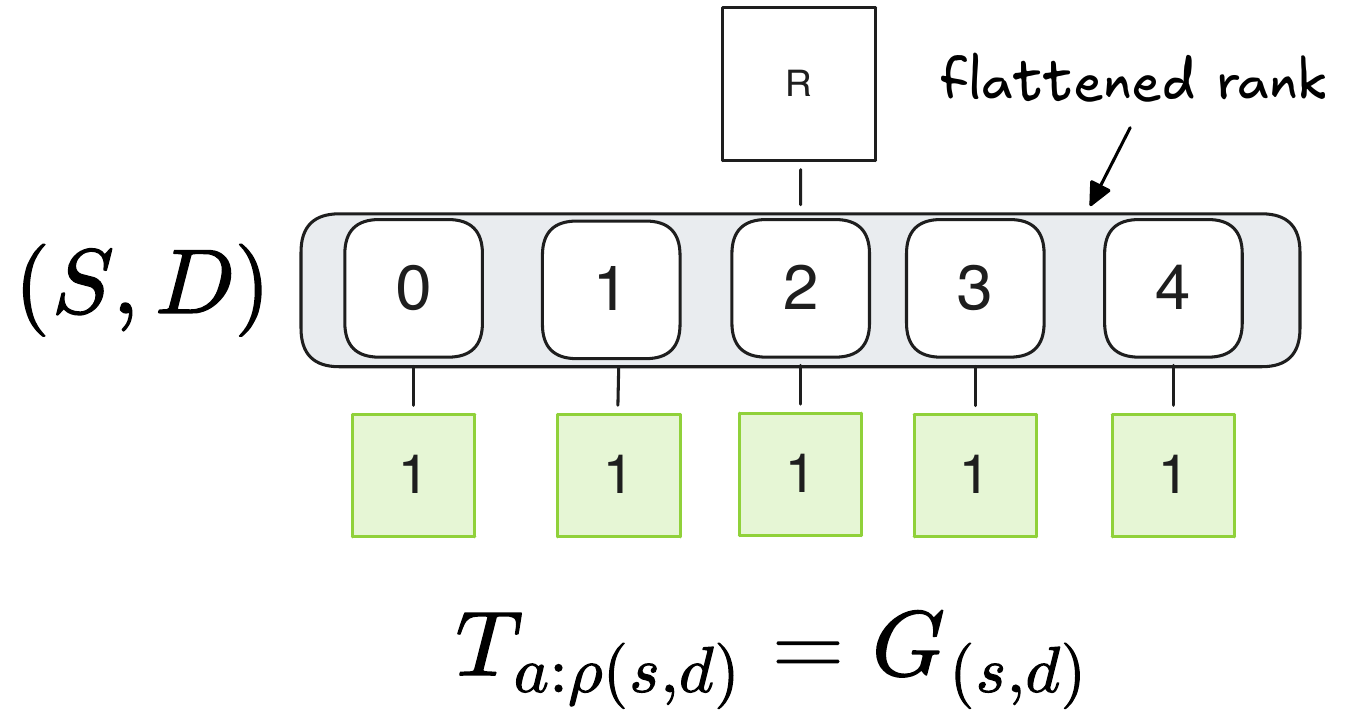}
    \caption{Position shifting.}
    \label{fig:position-right}
  \end{subfigure}

  \caption{Rank-variable rewrites: (a) Flattened rank and (b) Position shifting. The original graph is shown in Figure~\ref{fig:position-left}.}
  \label{fig:rewrites-group}
\end{figure}

The Einsum-Enabled Design Space (EEDS) builds upon the TeAAL pyramid of abstractions~\cite{Nayak:2023:TDF} (left side of Figure~\ref{fig:arch_constraints}) to structure the design space of graph algorithm implementations.
Traditional GPU graph optimization appears as a collection of loosely connected techniques, including load balancing strategies, frontier sorting, atomics, deduplication, and format choices (right side of Figure~\ref{fig:arch_constraints}).
EEDS organizes these techniques into four orthogonal concerns along the same axes as TeAAL.

Einsums enable separation of concerns, factoring complexity into smaller, composable transformations~\cite{Reade:1989:EFP, Goedicke:1990:PMS}.
We categorize these transformations into four levels: Einsum Space, Mapping Space, Format Space, and Code Space.
The first three are platform-independent; the final level captures platform-specific binding decisions.
Although architectural constraints may influence mapping or format choices (e.g., GPU-driven partitioning decisions), these influences can remain confined to specific levels of the abstraction.

Each point in EEDS corresponds to a concrete combination of choices across these four concerns.
Many well-known GPU optimizations emerge naturally as specific configurations within this structured space.
Using the BFS Einsum from Cascade~\eqref{eqn:edge_bfs}, we now enumerate these four levels of concern and examine what each implies in a graph algorithm context.

\subsection{(A) Einsum (EDGE) Space: Algebraic Manipulations}
\label{ssec:algebraic_manipulations}
Given an EDGE expression, one can apply algebraic manipulations to transform it into a different, but semantically equivalent, expression.
We term these \emph{algebraic manipulations}.
For example, FuseMax~\cite{Nayak:2024:FML_micro} proposes a specific ``pass-analysis'' manipulation on cascades that we consider a special case of algebraic manipulation.
Prior work in tensor algebra has not made algebraic manipulations a first-class concern. However, graph implementations highly benefit from exploring different algorithmic variants.
To this day, new graph algorithms are continually proposed to reduce the complexity of a given graph problem (e.g., Duan et al.~\cite{Duan:2025:BSB} or Liu and Tarjan~\cite{Liu:2022:SCC}).
We identify the following manipulations:
\textbf{A.1---Reassociate}: Within Einsums, we use parentheses to indicate which data values should be computed before other values are computed.
For example, the expression $F_{i+1, d} = (G_{s, d} \cdot F_{i, s})_{i, s, d} \cdot \, \neg P_{d}$ indicates that for a given $s,d$ point in the iteration space, the values corresponding to $G$ and $F$ should be computed first, then the result merged with the corresponding value in $P$ \footnote{Here, the $\cdot$ symbol is shorthand for the map/reduce actions in Equation~\eqref{seqn:advance}.}
In graph terminology, this is simply the step where active vertices collect their neighbors.
Reassociation reorders the parentheses.
The expression may now appear as $F_{i+1, d} = G_{s, d} \cdot (F_{i, s} \cdot \, \neg P_{d})_{i, s, d}$, where, for a given point in the iteration space, we now select the corresponding $F$ term and the corresponding $P$ term before merging both with $G$\@.
One can view this as a cross product between the $F$ and $P$ tensors.
In graph terminology, this is equivalent to creating a temporary edge between a source in the frontier and a node in the visited set, then checking if that edge exists in the graph.
The end result of either computation is the same, but the steps differ significantly.

\textbf{A.2---Commute}: If the scalar computation allows it, tensors can be commuted.

\textbf{A.3---Substitution} places one Einsum expression into another when portions of the expression are equivalent.

\textbf{A.4---Decomposition} decomposes an Einsum expression into separate Einsum expressions.
Decomposition explicitly creates new, intermediate tensors. We can decompose Equation~\eqref{seqn:advance} into two Einsums:
\begin{subequations}
    \begin{align}
    T_{i, s, d} &= G_{s, d} \langle \frac{+(\cap)_s}{^.} \rangle F_{i, s} \\
    F_{i+1, d} &= T_{i, s, d} \,\langle \frac{\leftarrow\!(\cap)_d}{{\text{ANY}(\cup)}_{s}} \rangle \, \neg P_{i, d},
    \end{align}
\end{subequations}
where $T$ is a sub-graph of the original graph that contains only those edges that had sources in the frontier.
Decomposition separates the computation into several phases that can potentially be mapped to task workers (see \textbf{A.8}).

\textbf{A.5---Uniform Partitioning} refers to tiling the iteration space or data into blocks, usually to enable data reuse.
Odemuyiwa et al.\ highlight a taxonomy to describe the space of tiling options~\cite{Odemuyiwa:2023:ASD}.
Uniform coordinate-space partitioning~\cite{Odemuyiwa:2023:ASD} evenly divides the coordinates of a particular rank into groups. For example, we can split the $s$ rank of Equation~\eqref{seqn:advance}:
\begin{equation}
    F_{i+1, d} = (G_{s_1, s_0, d} \cdot F_{i, s_1, s_0})_{d} \cdot \, \neg P_{d}
\end{equation}

Here, $s_1$ is the number of $s$ blocks, and $s_0$ is an $s$ coordinate within that block.
Suppose $s_0$ is 32---that is, there are 32 $s$ coordinates per block.
Then this expression is equivalent to grouping source vertices with ids between (0--31) into a group, vertices with ids between (32--63) into a group, and so on.

\textbf{A.6---Flattening} We can flatten two or more ranks into a single rank~\cite{Nayak:2023:TDF}.
Flattening is a transformation on the \emph{indices} of an Einsum expression.
Flattening ranks $s$ and $d$ of the graph, for example, would be expressed as:
    \begin{compactalign}
    G_{s, d} \rightarrow_\textit{flatten}G_{(s,d)}
    \end{compactalign}
where the tuple $(s,d)$ represents the newly flattened rank.
Flattening is often useful when followed by partitioning.
By first flattening, the partitioning step can now group disparate ranks together.
Figure~\ref{fig:position-middle} shows an example flattening.

\textbf{A.7---Convergence Manipulations}
The convergence condition ($\diamond: $) is an expression that returns \verb|true| or \verb|false| to determine when iteration should stop.
By using the manipulations above and other algebraic transformations, we can change the convergence condition (see variations 15--18 in Table~\ref{tab:variations}).

\textbf{A.8---Einsum/DAG Reordering}
An Einsum cascade forms a DAG of expressions~\cite{Nayak:2023:TDF}.
The DAG can be flattened, reordered, and serialized in different ways, while respecting dependencies.

\textbf{A.9---Rank Variable Rewrites}
Expressions in subscripts may be algebraically manipulated. For example, the generative rank $i$ in Equation~\eqref{seqn:advance} can be rewritten as $i-1$ in the input frontier and $i$ in the output frontier.
We introduce a new rank-variable rewrite that transforms the coordinates of non-empty points to represent their \emph{positions}. Given an ordered list of non-empty values, the position of each value is its index in that list.
\textbf{We extend EDGE} with an operator $\rho$ that rewrites a rank to range over these positions.

Figures~\ref{fig:position-left} and~\ref{fig:position-right} illustrate the transformation.
If $(s,d)$ contains five non-empty values, applying this transformation produces a rank ranging over $[0,5)$ with shape five: \(a : \rho(s,d)\),
where $a$ is the rewritten rank and $\rho$ maps each non-empty coordinate to its positional index.
This positional rewrite enables enqueue and dequeue semantics directly within the Einsum formulation (\S~\ref{sssec:gunrock_einsum}).

\subsection{(B) Mapping Space}
\label{ssec:space_time}
Space-time mappings impose a traversal order on the iteration space of the
Einsum, usually through a series of loop nests indicating when to walk each rank.
They also indicate which points in the space should be executed in parallel, and which should be executed sequentially.

\emph{Mapping} specifies \emph{how} to traverse the iteration and data space~\cite{Nayak:2023:TDF,Parashar:2019:TSA}.
Work from the tensor algebra community highlights different mapping transformations~\cite{Nayak:2023:TDF, Parashar:2019:TSA} that are also relevant in the graph world.

We highlight four common, but powerful, mapping choices:

\textbf{B.1---Traversal order} specifies, when implemented as a loop nest, the order in which to traverse the iteration space of the Einsum.

\textbf{B.2---Loop fusion and B.3---Loop fission} specifies whether to merge computation within one loop nest with computation in another loop nest.
For example, choosing to execute Equation~\eqref{seqn:advance} and Equation~\eqref{seqn:update} at the same time produces the two outputs $F_{i+1, d}$ and $P_{i+1, d}$ simultaneously.
This is equivalent to discovering a new node in the frontier and immediately updating its visited status.
Loop fission splits computation in a loop nest into multiple loop nests.

\textbf{B.4---Rank-swizzling} is a \emph{data-space} transformation, and involves changing the \emph{stored} order of the ranks of a tensor~\cite{Nayak:2023:TDF}.
If $G_{s, d}$ has an $s$-major layout in memory, then a rank-swizzle transforms it into a $d$-major layout.
In implementations, this can involve changing the data format (e.g., CSR to CSC), sorting a row-major COO structure to a column-major order, or sorting a list containing duplicates.

This transformation is often needed when removing duplicates from the active vertex set.
Suppose we want to represent a BFS algorithm that allows duplicates in the frontier across several iterations~\cite{Brahmakshatriya:2021:CGA,Wang:2017:GGG}.
This adds a new rank, $R$, to the $F$ tensor.
Collecting neighbors is now $G_{s, d} \cdot F_{i, r, s}$.
The extra $R$ rank corresponds to the \emph{position} of a vertex, $s$, in the frontier, to allow for duplicates to be represented in the Einsum.
The $R$ rank is dense, that is, it is non-empty for values of $r \in [0, R)$.
We assert that it can only point to a single $s$ coordinate.
Thus, duplicate sources can now appear in the input frontier, with the corresponding $r$ value indicating its position in the frontier.
For a given $i$ iteration, an $R$-major layout corresponds to an unsorted set of vertices.
This will lead to random accesses of the neighbor lists in the graph, as multiple, non-contiguous $r$ values may contain the same $s$ value.
Rank-swizzling $F$ such that it now has an order of $i\rightarrow s\rightarrow r$ means that duplicate $s$ vertices are now contiguous (see Variants~5--8 in Table~\ref{tab:variations}).
A low-level implementation (\S~\ref{ssec:codespace}) enables this by sorting the frontier.

\begin{figure}[t]
  \centering
  \begin{subfigure}[t]{0.48\columnwidth}
    \centering
    \includegraphics[width=\linewidth]{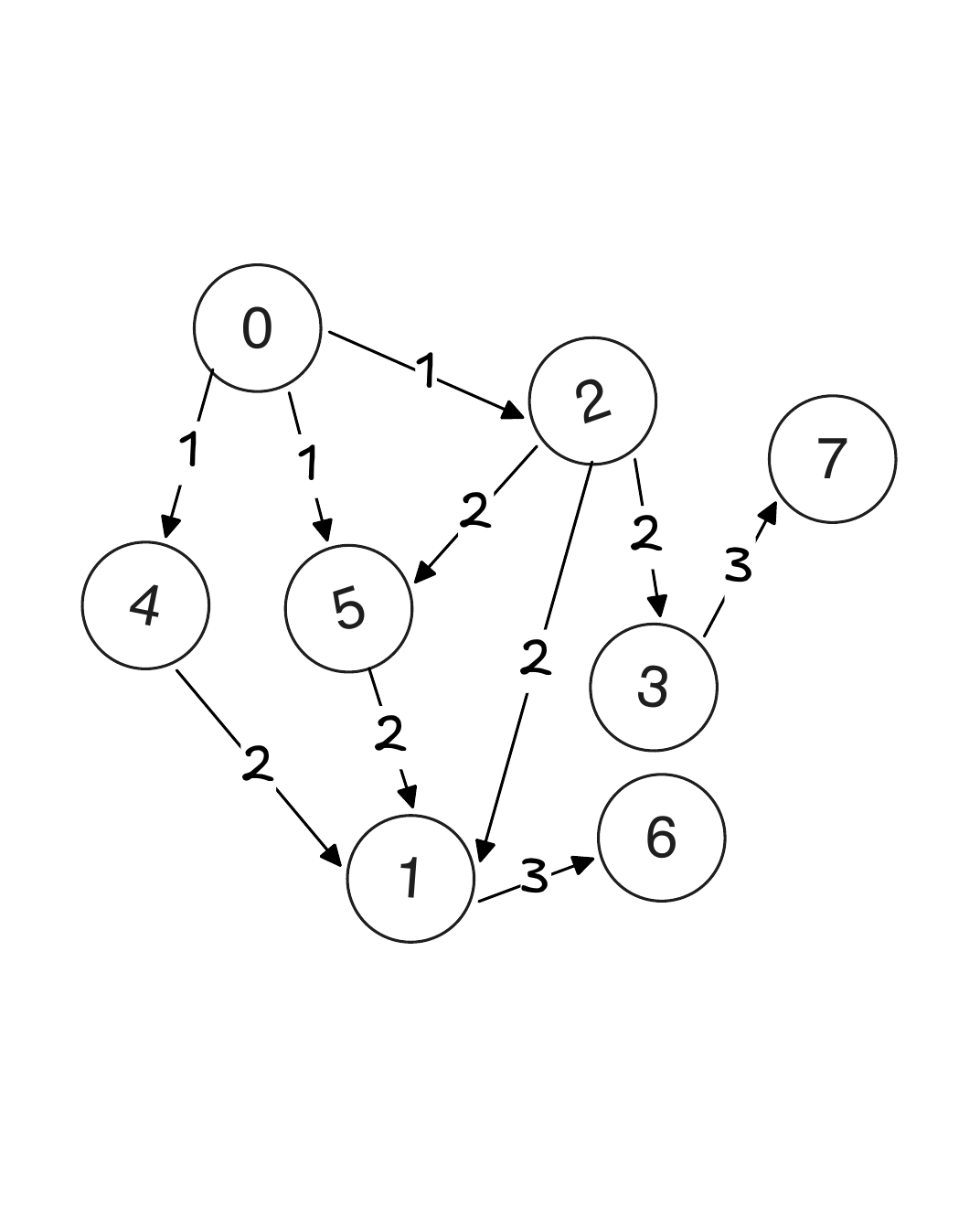}
    \caption{Weighted graph $G$.}
    \label{fig:eval-a}
  \end{subfigure}
  \hfill
  \begin{subfigure}[t]{0.48\columnwidth}
    \centering
    \includegraphics[width=\linewidth]{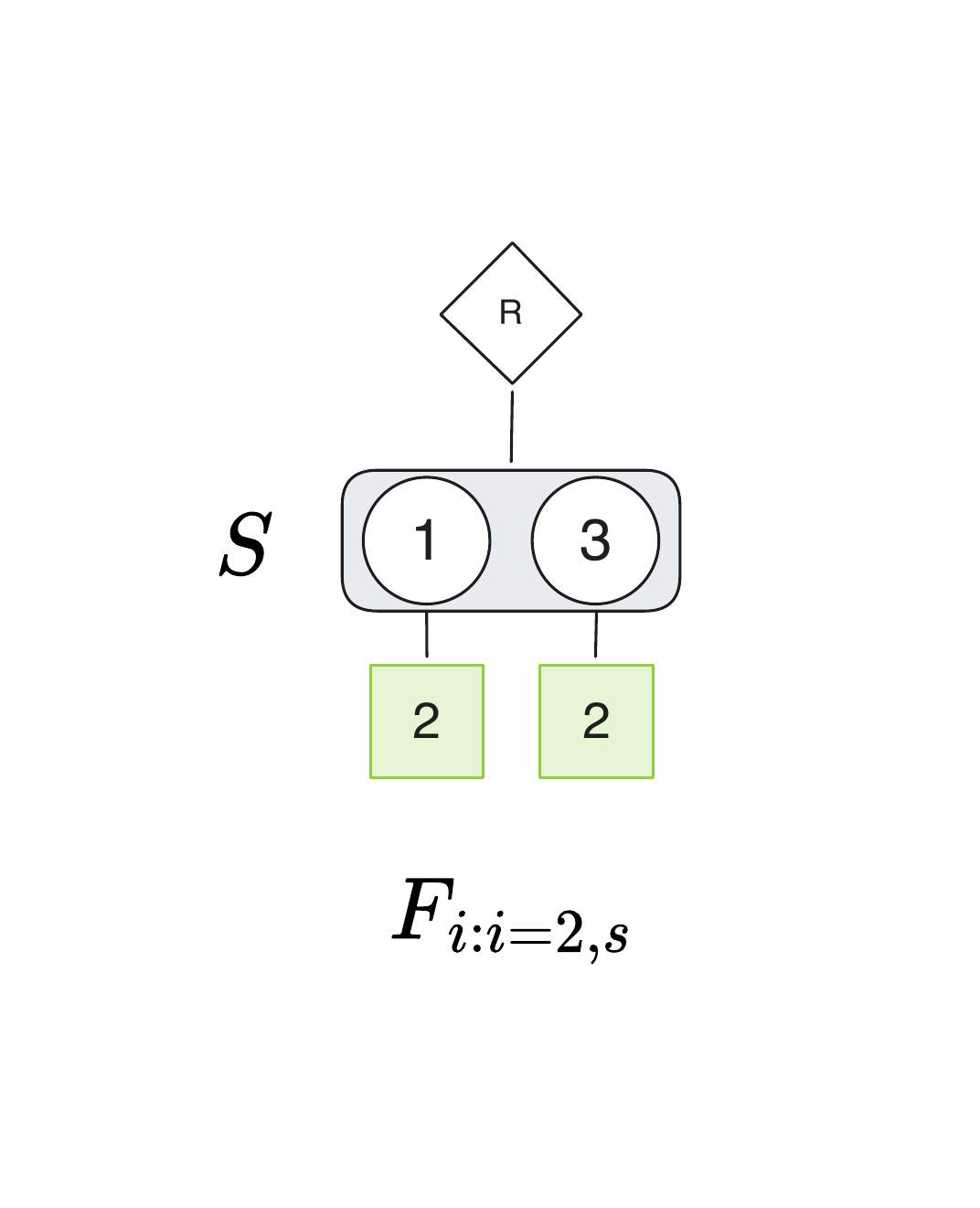}
    \caption{Frontier at iteration $i=2$.}
    \label{fig:eval-b}
  \end{subfigure}

  \begin{subfigure}[t]{0.48\columnwidth}
    \centering
    \includegraphics[width=\linewidth]{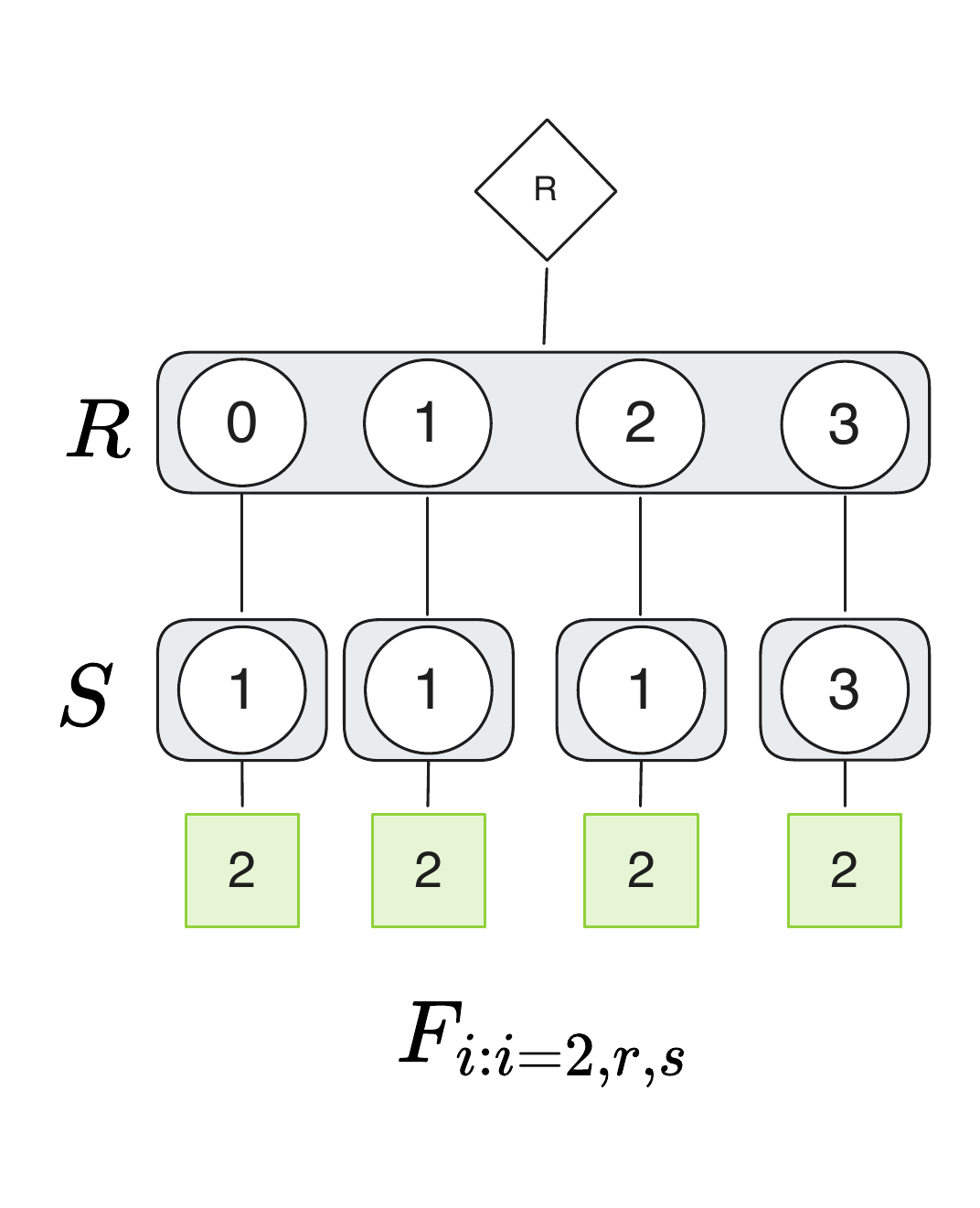}
    \caption{Allowing duplicates, introducing rank $R$.}
    \label{fig:eval-c}
  \end{subfigure}
  \hfill
  \begin{subfigure}[t]{0.48\columnwidth}
    \centering
    \includegraphics[width=\linewidth]{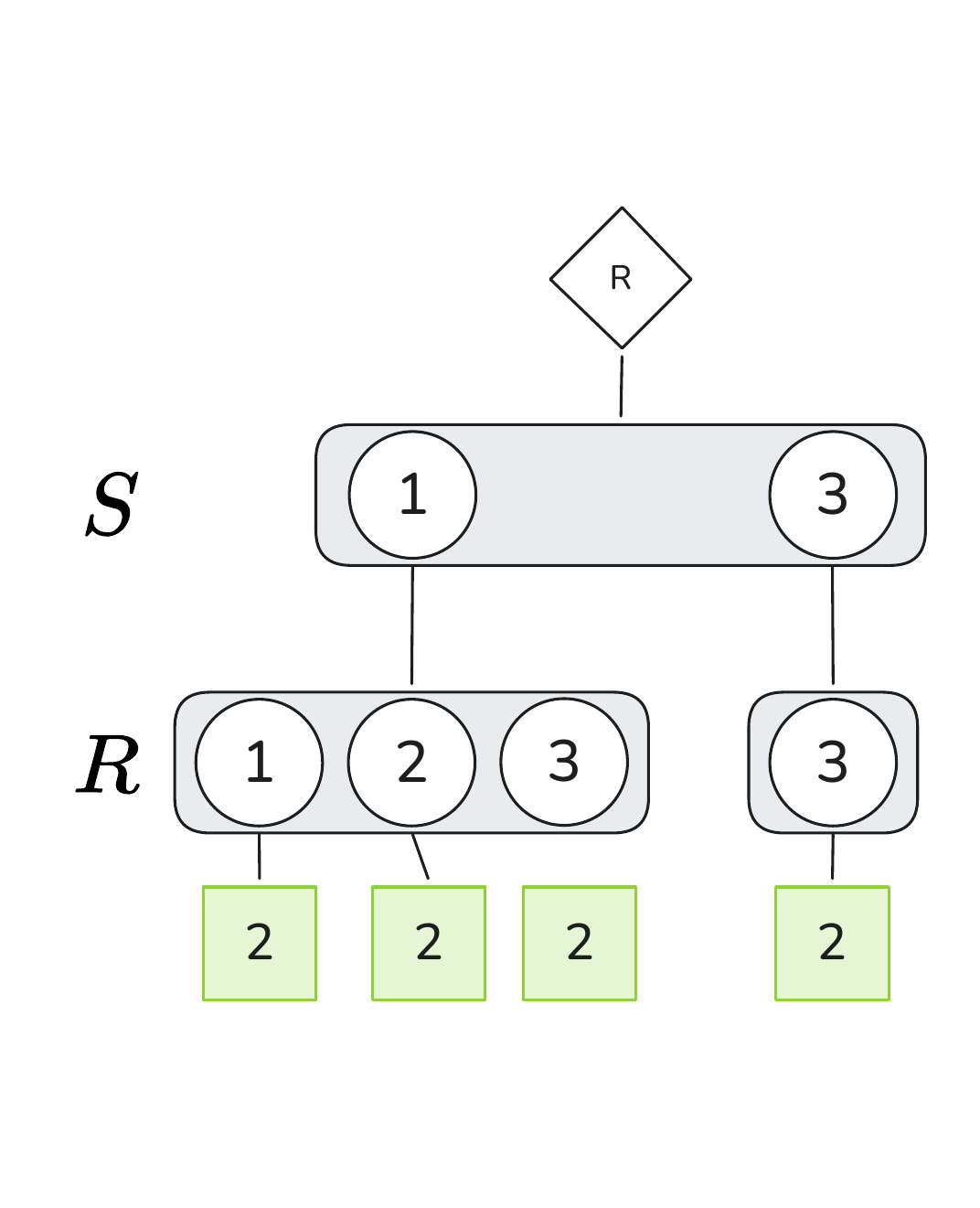}
    \caption{Rank swizzling (sorting in implementation).}
    \label{fig:eval-d}
  \end{subfigure}

  \begin{subfigure}[t]{0.48\columnwidth}
    \centering
    \includegraphics[width=\linewidth]{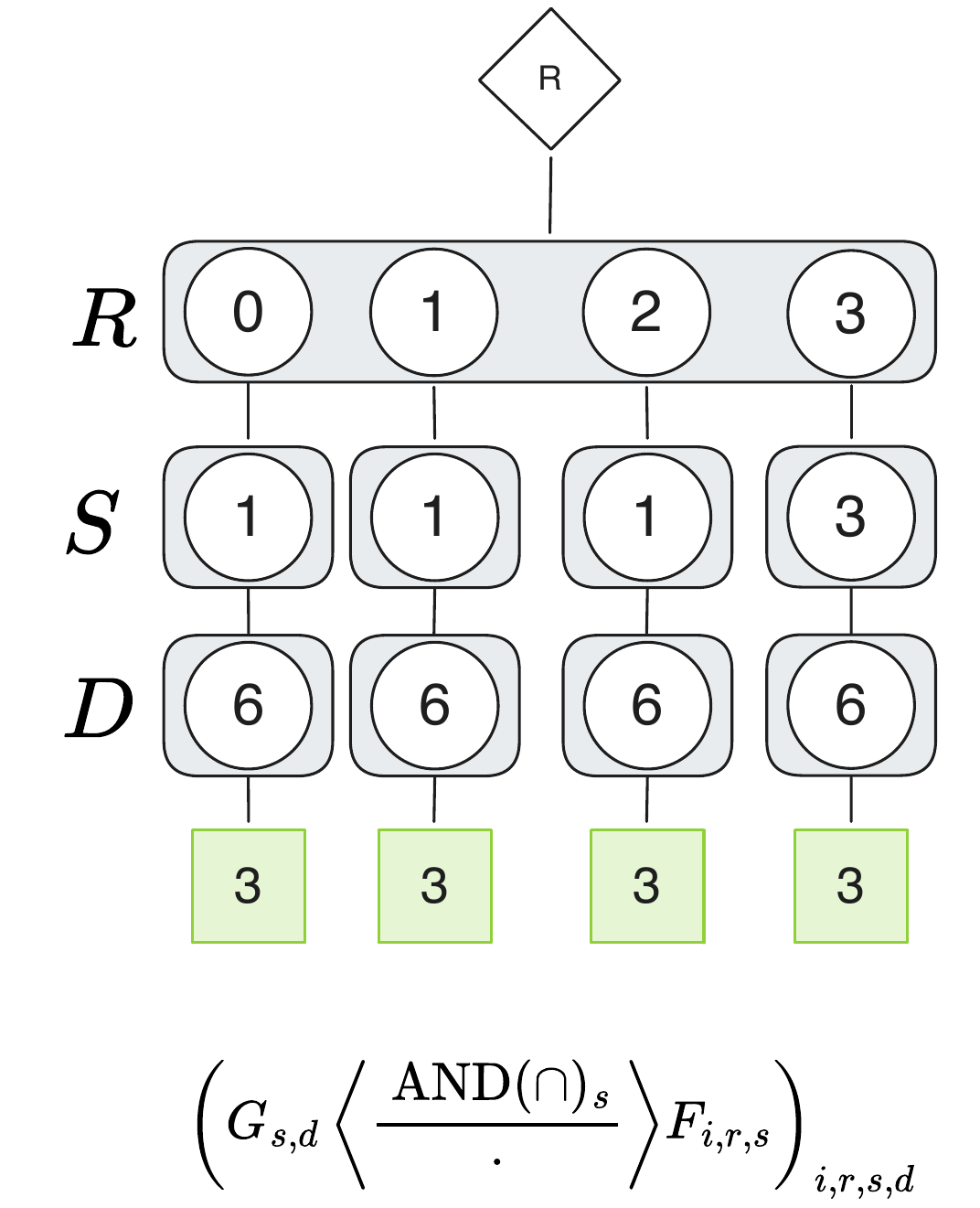}
    \caption{Temporary tensor after an advance on the graph.}
    \label{fig:eval-e}
  \end{subfigure}
  \hfill
  \begin{subfigure}[t]{0.48\columnwidth}
    \centering
    \includegraphics[width=\linewidth]{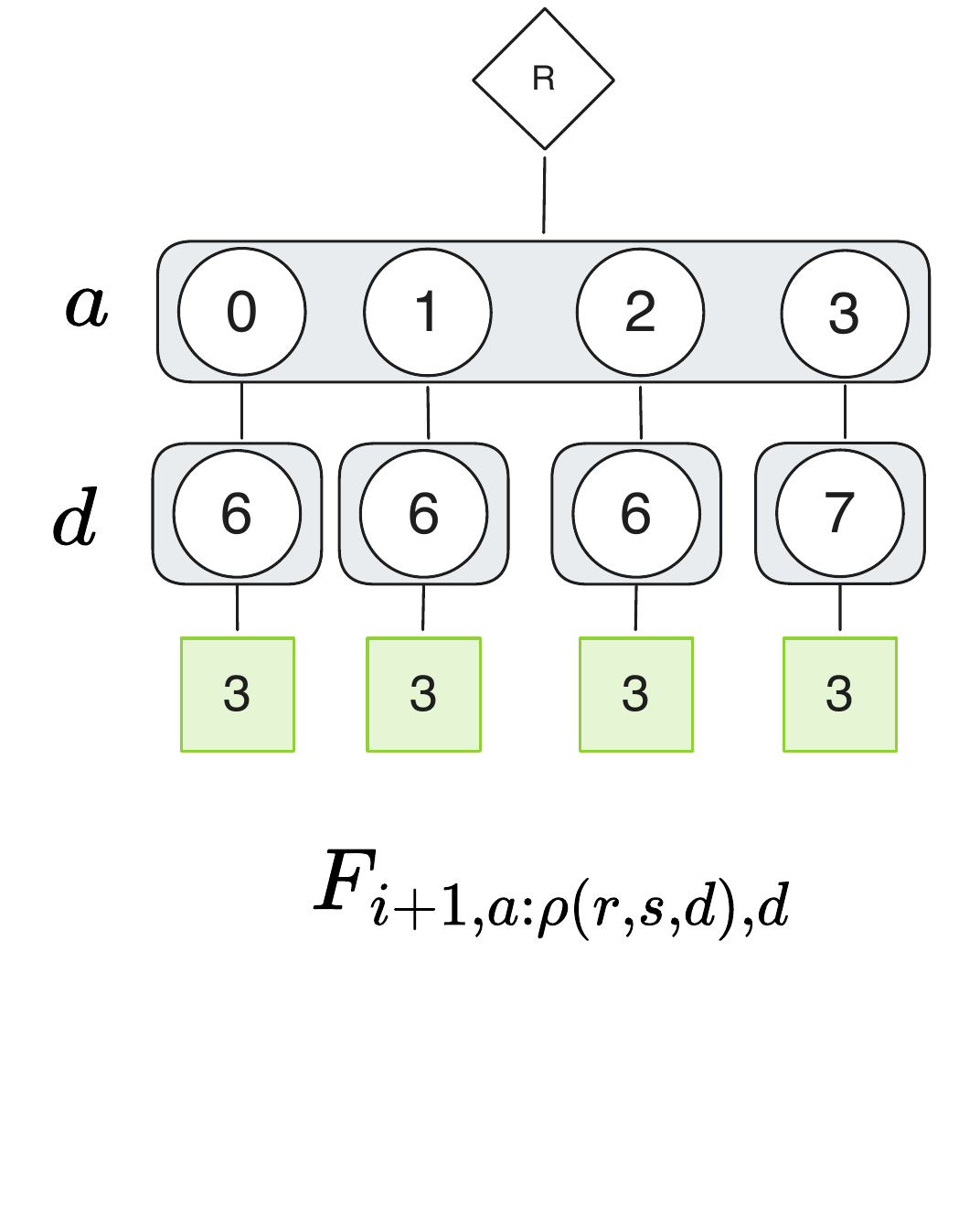}
    \caption{Resulting frontier representation after position shifting.}
    \label{fig:eval-f}
  \end{subfigure}
  \caption{A walkthrough of allowing duplicates ($R$ rank), rank swizzling, and position shifting ($a$ rank) on a small weighted graph.}
  \label{fig:eval-graphs}
  \vspace{-0.4em}
\end{figure}

Figure~\ref{fig:eval-graphs} walks through this sequence of transformations on a small weighted graph $G$ (Figure~\ref{fig:eval-a}): we form the frontier at iteration $i=2$ (Figure~\ref{fig:eval-b}), allow duplicates by introducing the $R$ rank (Figure~\ref{fig:eval-c}), apply rank swizzling, which manifests as sorting in an implementation (Figure~\ref{fig:eval-d}), produce a temporary tensor after an advance (Figure~\ref{fig:eval-e}), and finally apply position shifting, which introduces the $a$ rank (Figure~\ref{fig:eval-f}).

\textbf{B.5---Logical Scheduling} determines which points in the iteration space should be executed sequentially and which should be executed in parallel.

\subsection{(C) Format Space}\label{ssec:format}

Chou et al.~\cite{Chou:2018:FAF} highlight a format abstraction and corresponding costs, with subsequent work further exploring this space~\cite{Chou:2018:FAF,Wu:2022:SAA,Nayak:2023:TDF}.
In this work, we constrain implementations to the CSR format for graphs, and vector queues or boolean arrays (boolmap) for the frontier and visited set.

\subsection{(D) Code Space: Low-level Implementation}\label{ssec:codespace}

After selecting a point in the platform-independent optimization space for the above three concerns, we must now generate (either manually or automatically) platform-specific code.
In fact, some optimizations are only visible once we reach the implementation level (i.e., writing code!).
The Sparse Abstract Machine (SAM)~\cite{Hsu:2023:SAM} provides a set of ``binding'' primitives for sparse tensor algebra. Each primitive can be implemented in code space as needed. We specifically implement the ``intersecter'', ``unionizer,'' and ``swizzler'' (as various sorting methods).
This involves choosing from various software primitives, including (for instance), leader-follower intersection vs.\ double-sided intersection~\cite{Wu:2022:SAA}, different sorting algorithms~\cite{Cole:1988:PMS,CLRS:2009:ITA, Davidson:2012:EPM, Green:2012:GMP}, and different parallelization strategies~\cite{Dalton:2015:OSM,Osama:2023:PMG}.

\section{An Example Design-Space Exploration: From Push to Pull BFS}\label{ssec:pushpull}
Given a design choice at each level of concern in the EEDS, we can chain them together to produce an implementation point.
Insight into the best implementation strategies can be gained simply by looking at the Einsum and exploring the space of high-level mapping options.
As an example, we consider a common alternative to BFS\@.

Top-down, or ``push'' BFS, is the traditional implementation of BFS, and the focal point of this paper.
Push-BFS uses the sources in the frontier to determine which neighbors to include in the next frontier.
It took several decades, however, before the graph community discovered an alternate bottom-up, or ``pull'', BFS approach~\cite{Beamer:2012:DBS}; not even the existence of a matrix-based push-BFS formulation (which dates back to at least 2006~\cite{Gilbert:2006:HGA}) enabled this discovery.
Pull-BFS looks at the set of unvisited nodes, checks if any of their parents are in the frontier, and adds unvisited nodes that fit that criteria into the new frontier.
It is particularly useful in iterations where the frontier is more dense than the set of unvisited nodes.
We claim that a design-space exploration from a push-BFS Einsum expression, using the transformations we outline above, would have enabled such a discovery.

Push-BFS is simply a loop transformation mapping of a loop order of $s \rightarrow d$ applied to Cascade~\eqref{eqn:edge_bfs}.
Given the Einsum in Cascade~\eqref{eqn:edge_bfs}, we can now apply a series of algebraic manipulations, followed by loop transformation mappings, to go from a push-based formulation of BFS to pull-BFS\@:

\par{\textbf{Reassociate (A.1)}}: In graph terms, we combine the frontier and all unvisited nodes into a tensor where the set of unvisited nodes ($\neg P$) is repeated for each vertex in the frontier (intersection between $F$ and $\neg P$).
  We then perform an element-wise intersect on this temporary tensor with the graph, on the $s$ rank, to retrieve the neighbor list of vertices in the frontier, and on the $d$ rank to filter out visited nodes.
  \begin{compactalign}
    \label{eqn:rbfs_reassociate}
    F_{i+1, d} = G_{s, d} \,\langle \frac{+(\cap)_s}{{\textit{ANY}}(\cup)_{s}} \rangle\, \biggr(F_{i, s} \,\langle \frac{\leftarrow\!(\cap)_{s,d}}{^.} \rangle\, \neg P_{i, d}\biggl)_{i, s, d}
  \end{compactalign}

  \par{\textbf{Commute \ldots (A.2)}}: We then algebraically commute $P$ and $F$. This changes the $\leftarrow$ compute operator to a $\rightarrow$ operator, since at the graph level, we are interested in the values of the $F$ tensor (the current depths).
  \begin{compactalign}
    \label{eqn:rbfs_commute}
    F_{i+1, d} = G_{s, d} \,\langle \frac{+(\cap)_s}{{\textit{ANY}}(\cup)_{s}} \rangle\, \biggr(\neg P_{i, d} \,\langle \frac{\rightarrow\!(\cap)_{s,d}}{^.} \rangle\, F_{i, s}\biggl)_{i, s, d}
  \end{compactalign}
\par{\ldots then \textbf{Reassociate (A.1)} again}: After commuting $F$ and $P$ from the previous expression, we intersect the graph $G$ with the set of unvisited nodes, $\neg P$, on the $d$ rank. This creates a temporary tensor with ranks $s$ and $d$. We view this tensor as a subgraph that contains the vertices in $\neg P$ and their corresponding source nodes. We then intersect this subgraph with the current frontier $F$.
  If a vertex $d$ contains a source node in the frontier, we include it in the new frontier ($F_{i+1, d}$).
  \begin{compactalign}
    \label{eqn:rbfs_pull}
    F_{i+1, d} = \biggr(G_{s, d} \,\langle \frac{\leftarrow\!(\cap)_d}{^.} \rangle\, \neg P_{i, d}\biggl)_{i, s, d} \,\langle \frac{+(\cap)_s}{{\textit{ANY}}(\cup)_{s}} \rangle\, F_{i, s}
  \end{compactalign}
  \par{\textbf{Loop transformation: loop order (B.1)}}: Finally, we can apply a loop-order constraint of $d \rightarrow s$ to the previous expression. In graph terms, we look at children ($d$) before their parents ($s$).
  This gives us Pull-BFS\@.

Thus, given the original BFS Einsum, we simply needed four steps of algebraic manipulations and loop transformations to derive pull-BFS\@.

Of course our Einsum abstraction and these transformations are uninteresting unless they (1) can recover interesting implementations found in prior work, (2) discover novel implementations, and (3) can deliver performance wins.
Our next sections show that exploring different transformations is systematic, and both (1) map to prior work and novel implementations and (2) can be profitable.

\section{The EEDS of a GPU Graph Framework}
\label{sec:essentials}

There are two starting points for exploring the implementation space: (1) Begin with an Einsum description of the problem then apply various transformations to generate different implementations or
(2) Since EDGE can represent any point implementation, begin with a low-level implementation then raise the abstraction to its Einsum specification.
From this specification, we can then follow the process in (1).
Thus, \textbf{given an implementation, we can determine its EDGE abstraction, then apply transformations to explore other points in the space.}
Note that this raised abstraction is simply a point in the space generated by exploring the ``original'' Einsum (from step (1)).
It is possible to explore the implementation space by applying composable optimizations.
Indeed, even manually, we generate over 90 experiments all representing a point in the implementation space of BFS\@.
\S~\ref{sec:performance} discusses the performance of a subset of these variations.

\paragraph{Gunrock BFS}
\label{ssec:gunrock_bfs}

Any framework can serve as the software binding platform for a given point in the Einsum-Enabled Design Space.
We choose to explore our optimizations within Gunrock Essentials~\cite{Osama:2022:EOP, Wang:2017:GGG}, a state-of-the-art GPU graph framework that provides user-friendly APIs that map nicely to the tensor algebra-based design space.
We add low-level primitives to the codebase as needed when points in the optimization space are not supported by the baseline Gunrock library.
Our exploration, however, is not tied to Gunrock or to the GPU\@.

\subsection{Gunrock BFS as an EDGE abstraction}
\label{sssec:gunrock_einsum}

The Einsum for Gunrock's BFS is shown in Cascade~\ref{eqn:gunrock_einsum}\footnote{See Table~\ref{tab:compute} for a description of the compute operators.}
.
\begin{cascade}[t]
  \caption{Gunrock BFS cascade}
  \label{eqn:gunrock_einsum}
  \centering
  \begin{minipage}{\columnwidth}
  \begin{mdframed}[linewidth=0.6pt,roundcorner=2pt,
                  innertopmargin=4pt,innerbottommargin=4pt,
                  innerleftmargin=6pt,innerrightmargin=6pt]

  \begingroup
  \setlength{\abovedisplayskip}{6pt}
  \setlength{\belowdisplayskip}{6pt}
  \setlength{\jot}{2pt}
  \begin{subequations}
  \begin{align}
      &\triangleright\text{Tensor Declarations} \notag\\
    G^{S\equiv|V|, D\equiv|V|} &\rightarrow  \text{Int, empty=}0 \label{2seqn:G}\\
    P, F^{I, R, S\equiv|V|} &\rightarrow \text{Bool, empty=False} \label{2seqn:F}\\
    Depths^{I, D\equiv|V|} &\rightarrow \text{Int, empty=}\infty \label{2seqn:D}\\
    &\triangleright\text{Extended Einsum} \notag\\
    &\eqcomment{Is the current depth greater than the iteration count?}
    P_{i, d} &= \textit{Depths}_{i,d} \opsfrac{>(\cup)_s}{^.} (i+1)_d \label{seqn:depth_check}\\
    &\eqcomment{Advance on the predicate}
    F_{i+1, a: \rho(r,s,d), d} &= \biggl(G_{s,d}\ \opsfrac{\text{AND}(\cap)_s}{^.}\ F_{i,r,s}\biggr) _{i, r, s, d}\notag \\
            &\ \ \ \ \opsfrac{\text{AND}(\cap)_d}{^.}\ P_{i,d} \label{seqn:gunrock_advance} \\
    &\eqcomment{Track the new depths}
    T_{i, d} &= \textit{Depths}_{i,d} \opsfrac{\min(\cup)_d}{^.} (i+1)_d \label{seqn:depth_set} \\
    &\eqcomment{Update depths}
    \textit{Depths}_{i+1, d} &= \textit{Depths}_{i,d} \opsfrac{\lll(\cup)_d}{^.}\ \notag \\
    &\eqcomment{Reducing over $a$ first, only one vertex should update depths}
            &\ \ \ \  \biggl(F_{i+1,a,d}\ \opsfrac{\rightarrow(\cap)_d}{\text{ANY}(\cup)_a}\ T_{i, d}\biggr)_{i, d} \label{seqn:depth_update} \\
    \diamond&: ||F_{i}|| \equiv 0 \label{seqn:convergence}
  \end{align}
  \end{subequations}
  \endgroup

  \end{mdframed}
  \end{minipage}
\end{cascade}

Given the initial input frontier, Gunrock \emph{advances}: it collects the neighbors of the active vertex set (Equation~\eqref{seqn:gunrock_advance}).
This advance is predicated on if the depth at a particular vertex is still greater than the current iteration count, represented by $(i+1)_d$ (Equation~\eqref{seqn:depth_check}). The input frontier, $F$, allows duplicate vertices by adding an extra $R$ rank (see B.4 in \S~\ref{ssec:space_time}).
The output frontier renames this $r$ rank to $a$, as it needs to resize this rank.
The expression $a: \rho(r,s,d)$ takes each vertex in the frontier and every corresponding neighbor that survived the advance computation and populates the output frontier.
Within Gunrock, lambdas can also update auxiliary data structures: in this case, the depths are atomically updated (Equations~\eqref{seqn:depth_set} and~\eqref{seqn:depth_update}).
Note that this is a low-level implementation choice. It appears in the Einsum as a reduction over the $a$ rank, ensuring that only one of the potentially many updates to a given vertex's depth is applied.
The special update $\lll$ computation in Equation~\eqref{seqn:depth_update} takes the second operand if it exists, otherwise it takes the value in $\textit{Depths}_i$.
Finally, Gunrock BFS continues this process of advancing on a condition until a convergence condition is reached (Equation~\eqref{seqn:convergence}).
Overall, the Einsum specifications neatly map to the following steps in Gunrock: (1) advance on a predicate, (2) update depths, and (3) continue till convergence.
Although the Gunrock developers did not take this approach, we can express the baseline implementation as a series of transformations and mapping constraints on this Einsum.
Row one in Table~\ref{tab:variations} summarizes the algorithmic transformations and mapping constraints on this cascade for the baseline Gunrock BFS implementation.

This Einsum is our starting point from which we manually explore different points in the optimization space.

\section{Evaluation}
\label{sec:performance}

  \begin{figure*}
    \centering
    \includegraphics[width=1.05\textwidth]{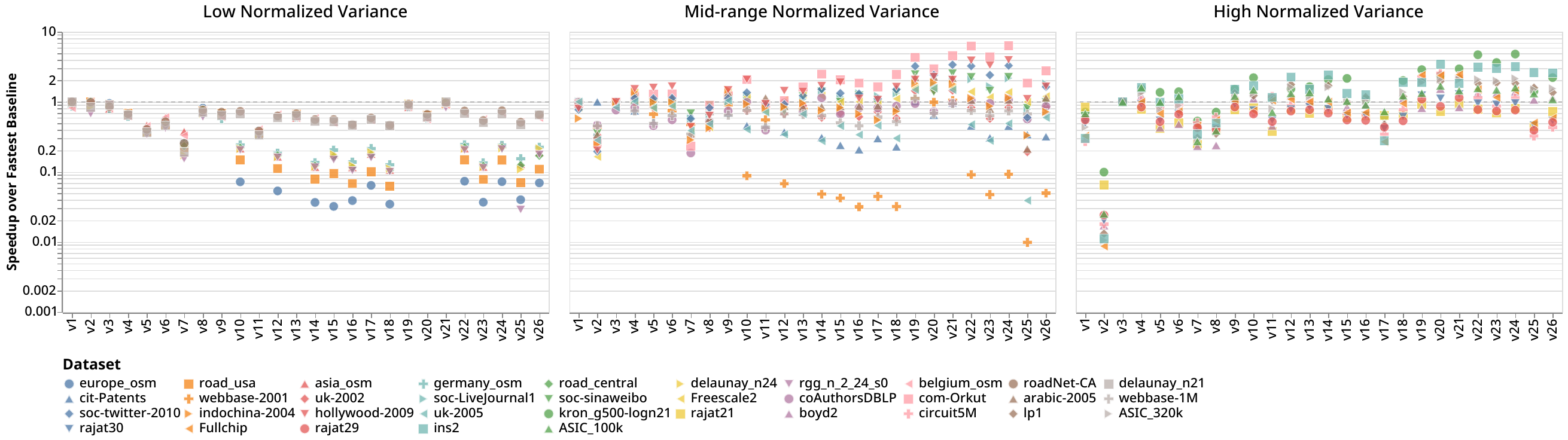}
    \caption{\textbf{Performance across variants.}
    Datasets are grouped by normalized degree variance (low: road-like; mid-range: social/scale-free; high: Kronecker and scientific/engineering graphs).
    Speedup is measured relative to the best baseline (variants~1--3).
    A per-dataset breakdown of best performance across all variants is shown in Figure~\ref{fig:small-multiples}.}\label{fig:summary-eval}
    \Description{Performance for different variants. Datasets are sorted in increasing normalized degree variance.
    Baseline implementations are variants 1--3. Speedup is determined over the best possible baseline variant.}
  \vspace{-0.4em}
  \end{figure*}

We now walk through how various well-defined Einsum-enabled transformations and mapping constraints correspond both to previously known BFS implementations
and to some new, potentially higher-performing variants.
Starting with the baseline Einsum in Cascade~\eqref{eqn:gunrock_einsum}, we apply chains of algebraic, mapping, format, and low-level transformations---as described in \S~\ref{sec:dse_with_ta}---to derive alternative BFS variants.
Using this approach, we were able to quickly discover over 90 implementation variations of BFS\@, categorized into 26 distinct variants.

To ensure an apples-to-apples comparison, we implement all variants within the Gunrock framework using Gunrock primitives (advance, filter), Thrust primitives~\cite{Bell:2011:TAP}, and ModernGPU’s \verb|merge path|~\cite{Baxter:2016:M2} where applicable.
Our objective is to demonstrate how an Einsum-enabled formulation enables systematic exploration of the design space, thus, we do not compare against external libraries. Instead, all variants are evaluated against one another and against the published Gunrock baseline~\cite{Osama:2022:EOP}, so that performance differences primarily reflect design choices rather than discrepancies across library implementations.
Implementing these variants in other libraries or on different hardware platforms (e.g., CPUs, alternative GPU architectures, or accelerators) may shift which points in the design space are optimal.
The variants presented here represent only a subset of the space unlocked by EEDS\@, illustrating both its expressive power and its potential to unlock new, high-performing implementations.

\subsection{Experiment Setup}
\textbf{Hardware and Software Environment.}
All experiments were conducted on a Linux workstation (Ubuntu 20.04 LTS) equipped with an NVIDIA Tesla V100 GPU (60GB memory) and four CPUs.
Code was developed using CUDA~12.1 and implemented within the Gunrock framework~\cite{Osama:2022:EOP}.

\textbf{Datasets.}
We evaluate our approach on 36 graph datasets drawn from the Stanford Network Repository~\cite{Leskovec:2014:SNAP}, the SuiteSparse Matrix Collection~\cite{Davis:2011:TUO, Kolodziej:2019:SMC}, and the Network Data Repository~\cite{Rossi:2013:NR}.
These datasets include both directed and undirected graphs ranging from 50K to 120M vertices ($|V|$) and approximately 200K to 1B edges ($|E|$).
To characterize graph structure, we use the \emph{normalized degree variance} metric proposed by Smith and Escudero~\cite{Smith:2020:NDV} (abbreviated “normalized variance”), which is invariant to graph size and density.
Empirically, we observe that low normalized variance graphs typically correspond to road networks; mid-range variance graphs correspond to social and scale-free networks; and high normalized variance graphs tend to include Kronecker and scientific/engineering graphs (e.g., \texttt{rajat30}, \texttt{ASIC100K}).

\textbf{Preprocessing and Methodology.}
All graphs are reordered using Reverse Cuthill--McKee (RCM) sorting~\cite{George:1984:RCM,Cuthill:1969:RBS}.
Since different repositories apply varying vertex orderings, we apply RCM uniformly to all datasets to ensure a consistent ordering and an apples-to-apples comparison across inputs.
For each BFS implementation, we perform 64 runs, randomly selecting a source vertex with degree greater than one (excluding self-loops), following the methodology of Beamer et al.~\cite{Beamer:2012:DBS} and the Graph500 benchmark suite~\cite{Graph500}.
Timing begins after graph and frontier loading and measures execution across all BFS iterations.

\subsection{Einsum-Enabled Design Space of BFS}
We list the Einsum-based transformations applied to generate each BFS variant in Table~\ref{tab:variations}.
\textbf{This is a design win: each variant can be described by a systematic application of transformations and mapping constraints.}
Transformation concerns are identified by the labels (A--D) in \S~\ref{sec:dse_with_ta}, and correspond to algebraic manipulation, space-time mapping, data format choice, and low-level implementation choices.

The following sections highlight a few of the various variants we generated in the EEDS\@, categorized according to their optimization strategy in the GPU space.
Each variant corresponds to a specific sequence of EEDS transformations applied to the baseline Einsum and composed with other variants (see Table~\ref{tab:variations} for the full list of transformations).
Overall, low normalized-degree variance graphs benefit from simpler variants (baseline implementations), thus we focus on mid-range and high normalized-degree variance graphs, where we see more significant performance differences across variants.

\subsubsection{Load Balancing Strategies}\label{sssec:lb}
The first design axis corresponds to space--time mapping transformations (Category B in \S~\ref{sec:dse_with_ta}).
In the Einsum formulation, load balancing arises from how iteration ranks are partitioned, flattened, and scheduled (sequentially or in parallel) to GPU threads, blocks, and grids.
Each combination (of partition, flatten, schedule) correspond directly to different, well-known, GPU work-distribution policies~\cite{Merrill:2012:SGG,Osama:2022:EOP}.

\begin{figure}[t]
  \centering
  \includegraphics[width=\columnwidth]{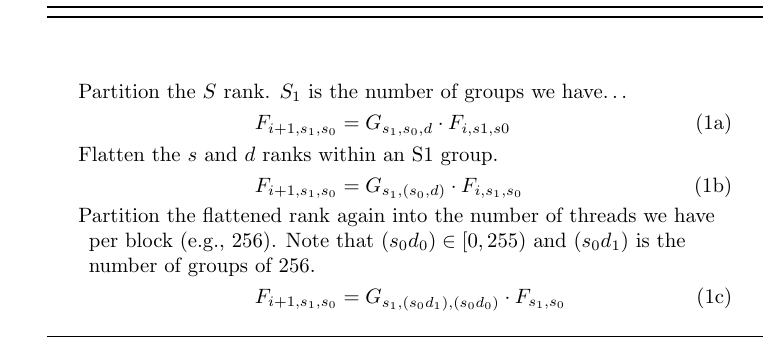}
  \caption{Block-mapped load balancing is a sequence of EEDS transformations.}
  \label{fig:lb_einsum_mapping}
  \vspace{-0.4em}
\end{figure}

Block-mapped (Variant 1) arises from partitioning, flattening, then partitioning again, with the second partitioning mapping to GPU blocks (see Figure~\ref{fig:lb_einsum_mapping}).
Thread-mapped load balancing (Variant 2) arises from partitioning the frontier into groups of threads, where each thread processes all neighbors of a single source vertex.
Merge-path (Variant 3) arises from flattening the edge frontier (flatten $(s, d)$), then partitioning total edge work evenly across threads using a merge-path strategy~\cite{Dalton:2015:OSM,Osama:2023:PMG}.
Gunrock already supports these three variants, and they are used as the baseline implementations in our evaluation.

\subsubsection{Work Pruning}\label{sssec:work_pruning}
Beyond load balancing, a key optimization region corresponds to transformations that prune redundant work.
In EEDS, this corresponds to reducing over intermediate ranks or altering the data representation to avoid materializing empty or duplicate entries.

\textbf{Removing Invalid Entries (Variant 4).}
In the baseline implementation, the output frontier allocates space for all neighbors of the input frontier, marking already visited vertices as invalid.
In Einsum terms, this corresponds to explicitly storing empty values in an uncompressed tensor.
An alternative is to remove invalid entries immediately via a \verb|filter| operation, producing a compressed frontier.
We observe that removing invalids is beneficial primarily for high normalized-degree variance graphs, where frontier imbalance is significant and many edges (pointing to the same destination) are filtered out.
For low-variance graphs, the additional filtering overhead often outweighs the benefit.

\textbf{Deduplication (Variants 5--8).}
Deduplication removes repeated vertices in the frontier, which in the Einsum view corresponds to reducing over a monotonically growing rank ($R$ rank in Cascade~\ref{eqn:gunrock_einsum}).
While deduplication is well known in prior work~\cite{Brahmakshatriya:2021:CGA,Merrill:2012:SGG}, the EEDS makes it explicit as a rank-reduction transformation.
Applied in isolation (Variant 5), deduplication provides little benefit due to the high cost of device-wide \emph{rank swizzling}, which corresponds to sorting.
However, when composed with invalid removal (Variant 4), it yields speedups for some graphs with mid- to high- normalized variance, where multiple sources point to the same destination.
Variants 6--8 explore device-wide frontier swizzling (i.e., sorting), swizzling within a frontier partition (block-level sorting), and deduplicating within a frontier partition (block-level deduplication).
Variant 6 successfully reduces the overhead of full deduplication in some datasets by placing duplicate vertices contiguously to minimize random accesses to global memory.

\textbf{Best-Effort Deduplication (Variant 9)}
This approach simply deletes contiguous vertices in the frontier without sorting.
This is equivalent to a localized rank-swizzle in the tensor.
Compared to the other deduplication variants, this approach consistently enables performance improvements for mid- to high- normalized variance graphs, without incurring the overhead of sorting.

\textbf{Boolmap Frontier (Variant 10).}
A dense boolmap frontier eliminates duplicates by construction.
This representation is beneficial when frontiers are dense, but incurs significant overhead when frontiers are sparse, as it requires instantiating storage proportional to $|V|$.

\subsubsection{Conflict and Predicate Management (Variants 11--14)}\label{sssec:conflict}

Another design axis concerns how update conflicts and predicates are handled during the advance step.

\textbf{Resolving Update Conflicts (Variants 11--12).}
When multiple sources attempt to update the same destination, atomic conflicts may arise.
To mitigate this, we separate the $\textit{Depths}$ array into read-only and write-only copies, consistent with the Einsum formulation, which requires only $\textit{Depths}_i$ and $\textit{Depths}_{i+1}$ (see \S~\ref{ssec:codespace}, ping-pong buffering).
This eliminates atomic updates during advance.

We observe improvements \emph{when combined with a boolmap frontier} (Variant 12), particularly in high normalized-degree variance graphs where many sources point to the same destination.
In these cases, reducing serialization outweighs the additional memory traffic and per-iteration copy cost.
Moreover, because the boolmap frontier prevents duplicate insertions by construction, there is no need for explicit deduplication (as with Variant 11, which performs poorly), further reducing overhead.

\textbf{Advance Before Updating Depths (Variants 13--14).}
In this variant, advance is performed without applying a predicate.
\emph{All} neighbors of frontier vertices are first written to a temporary frontier, and a subsequent \verb|filter| kernel updates the $\textit{Depths}$ array and removes invalid entries.
This restructuring reduces branch divergence in the advance kernel, since \emph{all} threads uniformly perform neighbor expansion.
The filtering step can then be executed with minimal divergence.
We compose this transformation with prior pruning strategies (invalid removal and deduplication) and select the best-performing configuration for evaluation.

\subsubsection{State Encoding with Visited Masks (Variants 19--26)}
A final design axis concerns how traversal state is encoded and used to prune redundant computation.

\textbf{Explicit Visited Mask (Variants 19--24).}
These variants introduce a separate visited array $P_d$ to prevent advancing on previously visited vertices.
During advance, neighbors are checked against $P_d$ before updating $\textit{Depths}$, reducing redundant edge traversals.
Variants differ along two dimensions: whether the visited update is fused with advance or performed in a subsequent \verb|filter| kernel, and whether atomic updates to $\textit{Depths}$ are removed (see Section~\ref{sssec:conflict}).
Across these configurations, we observe moderate performance improvements (geometric mean $0.9\times$--$1.5\times$), particularly for high normalized-degree variance graphs where redundant neighbor advance is common.

\textbf{Visited-as-Frontier (Variant 25).}
Rather than maintaining both a frontier and a visited array, this variant replaces the frontier with the visited array itself~\cite{Yang:2018:IPE}.
Although this eliminates duplicate tracking and avoids atomic depth updates, it causes previously visited vertices to be relaunched in every iteration.
Performance improvement depends on the relative cost of redundant launches versus maintaining separate data structures and atomics.

\textbf{Three-Array Visited Strategy (Variant 26).}
Building on the mask-based approach, this novel variant maintains three visited arrays: the current state, the previous iteration’s state, and the next state.
Rather than repeatedly launching the advance kernel on vertices that have already been visited, the current visited array only advances on a vertex if it was not present in the previous iteration.

\subsubsection{Key Highlights}
Overall, BFS performance is not dictated by a single optimization, but by the interaction between load balancing, work pruning, conflict management, and state encoding.
Low-variance graphs are well served by simpler implementations (see first chart in Figure~\ref{fig:summary-eval}), whereas high-variance graphs require pruning and contention reduction to avoid redundant work and serialization.
By expressing BFS as an Einsum, we expose these optimization optimizations explicitly and enable systematic traversal of the design space.
This formulation not only recovers well-known strategies, but also surfaces new, effective implementations, demonstrating the practical power of the Einsum-enabled design space.

\begin{figure*}
    \centering
    \includegraphics[width=\textwidth]{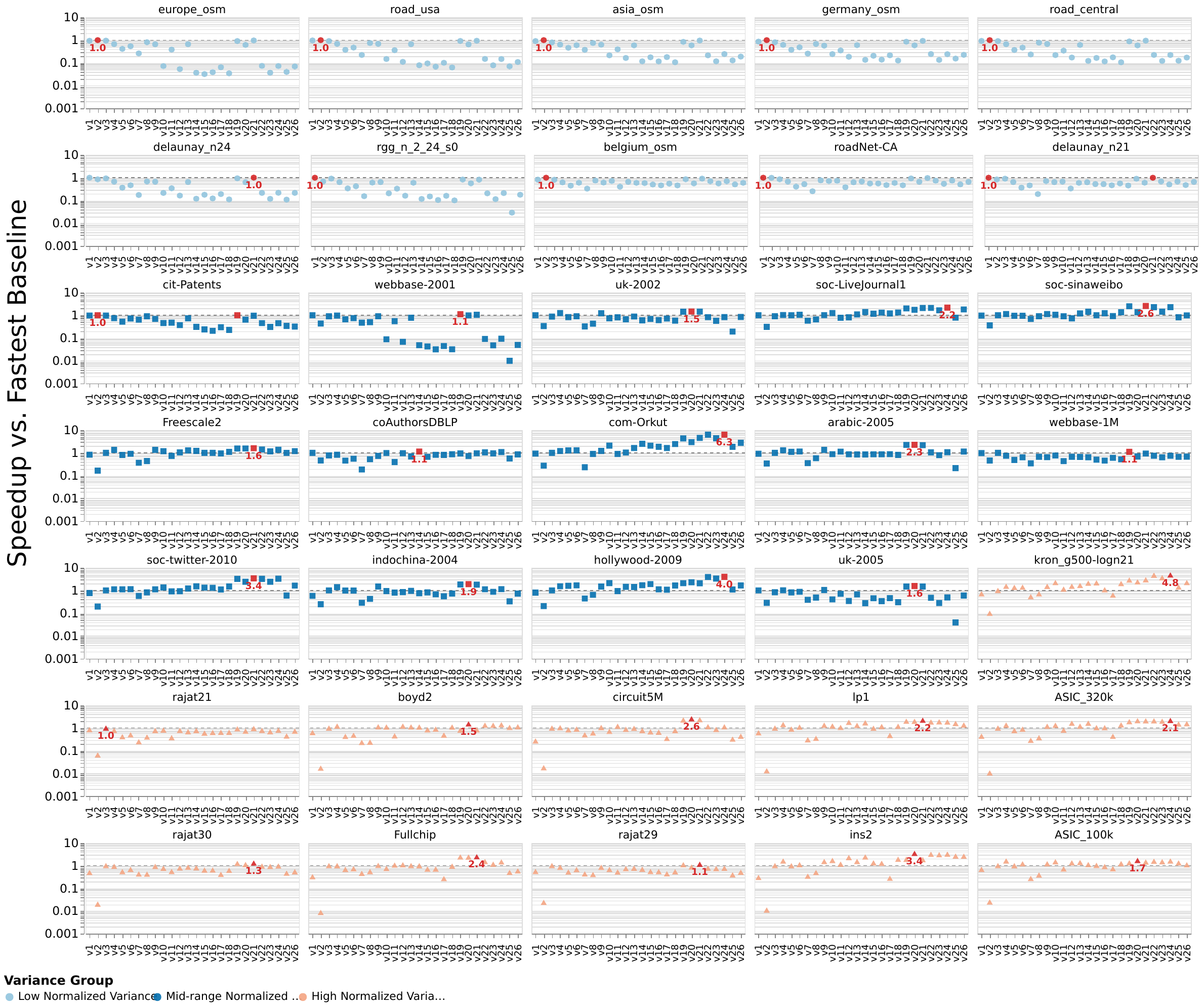}
    \caption{
        Per-dataset speedup of all 26 BFS variants.
        Datasets are ordered by increasing normalized degree variance.
        Speedup is reported relative to the fastest baseline (Variants~1--3).
        Highlighted (in red) are variants that achieve the best performance speedup for the given dataset.
    }\label{fig:small-multiples}
    \Description{Performance for different variants. Datasets are sorted in increasing normalized degree variance.
    Baseline implementations are variants 1--3. Speedup is determined over the best possible baseline variant.}
  \vspace{-0.4em}
  \end{figure*}

Figure~\ref{fig:small-multiples} shows the per-dataset performance of all 26 BFS variants. Datasets are ordered by increasing normalized degree variance, and speedup is reported relative to the fastest baseline.

\afterpage{%
\begin{landscape}
  \begin{table*}[]
  \caption{Summary of the 26 BFS variations we implemented in Gunrock. (A.x), (B.x), (C), and (D) refer to the labeled transformations and mapping constraints applied from \S~\ref{sec:dse_with_ta} that we applied to get from the ``baseline'' Einsum in Cascade~\eqref{eqn:gunrock_einsum} to that variation's implementation. E~\eqref{seqn:depth_check}, E~\eqref{seqn:gunrock_advance}, E~\eqref{seqn:depth_set}, and E~\eqref{seqn:depth_update} refer to Equations~\eqref{seqn:depth_check}, \eqref{seqn:gunrock_advance}, \eqref{seqn:depth_set}, and~\eqref{seqn:depth_update}, respectively, in the baseline Einsum. Variants marked with * indicate novel BFS implementations discovered in this work.
  }
  \label{tab:variations}

  \begingroup
  \fontsize{9.5pt}{11.2pt}\selectfont
  \setlength{\tabcolsep}{3pt}
  \renewcommand{\arraystretch}{1.02}

  \begin{adjustbox}{max width=1.05\textwidth}
  \begin{tabular}{cll}
  \toprule
  \multicolumn{1}{c}{\textbf{Variation}} &
    \textbf{EDGE Transformations Applied} &
    \textbf{Corresponding Graph Implementation} \\ \midrule
  1 &
    \begin{tabular}[c]{@{}l@{}}
      (B.1):  $I \rightarrow A \rightarrow R \rightarrow S \rightarrow D$\\
      (A.3): substitute  E~\eqref{seqn:depth_check} into E~\eqref{seqn:gunrock_advance} (produces only $F_{i+1}$)\\
      (A.3): substitute  E~\eqref{seqn:depth_set} into E~\eqref{seqn:depth_update} \\
      (B.2): fuse E~\eqref{seqn:gunrock_advance} with E~\eqref{seqn:depth_update}\\
      (A.5): partition on $R$ $\rightarrow_{generates} R_1, R_0$\\
      (A.6): flatten ranks $R_0, S, D$ $\rightarrow_{generates} (R_0, S, D)$\\
      (A.5): partition on $(R_0,S,D)$ $\rightarrow_{generates}  (R_0, S, D)_1, (R_0,S,D)_0$\\
      (B.5): parallel on $R_1$, $(R_0, S, D)_0$\\ (B): sequential on $(R_0, S, D)_1$\\ (C): frontier queue\end{tabular} &
    Applying Load Balancing (block) \\ \midrule

  2 &
    (A.5): partition on $R \rightarrow_{generates} R1, R0$ &
    Applying Load Balancing (thread~\cite{Osama:2022:EOP}) \\ \midrule

  3 &
    \begin{tabular}[c]{@{}l@{}}
      (A.6): flatten on $R,S,D \rightarrow_{generates} (R,S,D)$\\
      (A.5): partition on $(R,S,D)$\end{tabular} &
    Applying Load Balancing (merge~\cite{Osama:2022:EOP}) \\ \midrule

  5 &
    (A.4) decompose
    E~\eqref{seqn:depth_update} into
    E~\eqref{seqn:depth_update}.1 populate ($\lll$) on Depths, and
    E~\eqref{seqn:depth_update}.2 reduction over $a$ on $F_{i+1}$ &
    Deduplication \\ \midrule

  6* &
    (B.4): swizzle $F$ on ranks $R, S \rightarrow S, R$ &
    Frontier sorting \\ \midrule

  7* &
    \begin{tabular}[c]{@{}l@{}}
      (A.5, A.6, A.5): partition-flatten-partition (see variation 1)\\
      (B.4): swizzle \emph{within} an $(R_0,S,D)_1$ tile\end{tabular} &
    Block-level frontier sorting \\ \midrule

  8* &
    \begin{tabular}[c]{@{}l@{}}(A.5, A.6, A.5):
      partition-flatten-partition (see variation 1)\\
      (A): reduce over $a$ \emph{within} an $(R_O,S,D)_1$ tile\end{tabular} &
    Block-level deduplication \\ \midrule

  9 &
    (B.4):  localized rank-swizzling &
    Best-effort deduplication \\ \midrule

  10 &
    \begin{tabular}[c]{@{}l@{}}
      (A): Remove the $R/A$ rank and reduce over $S$ in  E~\eqref{seqn:depth_check}\\
    (C) + (D): Uncompressed boolmap frontier\\
    (A.7): $\diamond: T = F_{i, d}, T \equiv 0$ (i.e., reduce over $d$ and check if the result is 0)\end{tabular} &
    Boolmap frontier (instead of queue) \\ \midrule

  11* &
    \begin{tabular}[c]{@{}l@{}}
      (A.4): Compute E~\eqref{seqn:depth_set} and E~\eqref{seqn:depth_update} separately\\
      (C) + (D): Maintain $D_{i}$ and $D_{i+1}$ as separate, physical data structures\end{tabular} &
    Remove atomics (keep two copies of Depths array) \\ \midrule

  12* &
    Variation 10 + Variation 11 &
    Boolmap frontier + Remove atomics \\ \midrule

  13 &
    \begin{tabular}[c]{@{}l@{}}
      (A.4): Decompose E~\eqref{seqn:gunrock_advance} s.t.  E~\eqref{seqn:gunrock_advance}.1 $Temp = G \cdot F_i$ and E~\eqref{seqn:gunrock_advance}.2: $F_{i+1} = Temp \cdot P_i$\\
      (A.8): Reorder  E~\eqref{seqn:depth_check}, E~\eqref{seqn:gunrock_advance}.1, E~\eqref{seqn:gunrock_advance}.2 $\rightarrow$ E~\eqref{seqn:gunrock_advance}.1,  E~\eqref{seqn:depth_check}, E~\eqref{seqn:gunrock_advance}.2\\
      (B.2): Fuse  E~\eqref{seqn:depth_check} with E~\eqref{seqn:gunrock_advance}.2
    \end{tabular} &
    Advance before updating Depths array \\ \midrule

  14 &
    Variation 10 + Variation 13 &
    Boolmap frontier + Advance before updating Depths array \\ \midrule

  15 &
    \begin{tabular}[c]{@{}l@{}}Variation 10\\
      (A.7): $\diamond: F_{i+1} \equiv F_{i}$\end{tabular} &
    Converge by comparing frontiers \\ \midrule

  17 &
    (A.7) + (D): $\diamond:||F_{i+1}|| \equiv 0$ by checking for non-zeros in parallel &
    Check frontier occupancy using ``thrust::find'' \\ \midrule

  18 &
    Variation 12 + Variation 17 &
    Boolmap frontier + Remove atomics + Converge through occupancy check \\ \midrule
  19 &
    \begin{tabular}[c]{@{}l@{}}(A): Add generative rank to $T_d \rightarrow_{becomes} T_{i,d}$  s.t.  E~\eqref{seqn:depth_check}: $T_{i+1, d} = Depths\ldots$\\ (A.3): Substitute $T_{i}$ into E~\eqref{seqn:gunrock_advance} (apply previous iteration's $T$ as a mask)\\ (B.2): Fuse  E~\eqref{seqn:depth_check} and E~\eqref{seqn:gunrock_advance} and E~\eqref{seqn:depth_set} and E~\eqref{seqn:depth_update} (advance while updating the mask)\end{tabular} &
    Applying an explicit mask to edge accesses \\ \midrule

  20 &
    \begin{tabular}[c]{@{}l@{}}
      (A.8): Reorder such that  E~\eqref{seqn:depth_check} is now the last, unfused, step $\rightarrow$ E~\eqref{seqn:gunrock_advance}, E~\eqref{seqn:depth_set}, E~\eqref{seqn:depth_update}, E~\eqref{seqn:depth_check}\\
      (A.3): Substitute $P_{i-1}$ into E~\eqref{seqn:gunrock_advance} (apply previous iteration's $P$ as a mask)\\
      (B.2): fuse E~\eqref{seqn:gunrock_advance}, E~\eqref{seqn:depth_set}, and E~\eqref{seqn:depth_update} (advance while updating $Depths$)
    \end{tabular} &
    Applying an explicit mask to edge accesses + Updating mask in a separate kernel \\ \midrule

  21* &
    \begin{tabular}[c]{@{}l@{}}Starting point: Variation 19\\
      (A): Through induction, replace  E~\eqref{seqn:depth_check} with $P_{i+1, d} = P_{i, d} \langle \frac{\text{OR}_d}{^.} \rangle F_{i+1, d}$\\
      (A): Through induction (from previous step), replace $\min$ in E~\eqref{seqn:depth_set} with $(i+1)$ (i.e., update based on new frontier)\\
      (B.2): Fuse E~\eqref{seqn:depth_check}, E~\eqref{seqn:gunrock_advance}, E~\eqref{seqn:depth_set}, E~\eqref{seqn:depth_update} \\
      (A.8): Reorder to E~\eqref{seqn:gunrock_advance}, E~\eqref{seqn:depth_set}, E~\eqref{seqn:depth_update}, E~\eqref{seqn:depth_check}
    \end{tabular} &
    Applying an explicit mask to edge accesses + removing atomics and extra comparators \\ \midrule

  22 &
    Variation 19 + Variation 10 &
    Boolmap frontier + Applying an explicit mask to edge accesses \\ \midrule

  23 &
    Variation 20 + Variation 10 &
    Boolmap frontier + Applying an explicit mask to edge accesses + Updating mask in a separate kernel \\ \midrule

  24* &
    Variation 21 + Variation 10 &
    Boolmap frontier + Applying an explicit mask to edge accesses + removing atomics and extra comparators \\ \midrule

  25* &
    \begin{tabular}[c]{@{}l@{}}Starting with Variation 21:\\
      (A.3): By induction, substitute $F$ with $P$ in all sub-Einsums: $\rightarrow$ E~\eqref{seqn:gunrock_advance}: $P_{i+1, d} = G_{s, d} \cdot P_{i, s} \cdot P_{i,d}$\\
      (A): From previous step, delete  E~\eqref{seqn:depth_check} ($P_{i+1, d}$ is now generated at E~\eqref{seqn:gunrock_advance})\\
      (B.2): Fuse E~\eqref{seqn:gunrock_advance} and E~\eqref{seqn:depth_set} and E~\eqref{seqn:depth_update}\end{tabular} &
    Use visited array ($P_i$) as frontier AND mask \\ \midrule

  26* &
    \begin{tabular}[c]{@{}l@{}}Starting with Variation 21:\\
      (A.3) Through substitution, note that $F_{i, d} = P_{i, d} \neg P_{i-1, d}$\\
      (A.3) Substitute expression from previous step into E~\eqref{seqn:gunrock_advance} $\rightarrow$ E~\eqref{seqn:gunrock_advance}: $F_{i+1, d} = G_{s, d} \cdot (P_{i, s} \cdot \neg P_{i-1, s}) \cdot P_{i, d}$\end{tabular} &
    Use visited array ($P_i$) as frontier AND mask + Check vertex is not in previous iteration's mask before advance \\ \bottomrule
  \end{tabular}%
  \end{adjustbox}
  \endgroup

  \vspace{-0.4em}
  \end{table*}
  \end{landscape}
}

\section{Conclusion}

We presented the Einsum-Enabled Design Space (EEDS), a structured abstraction for reasoning about graph algorithm implementations. By expressing breadth-first search (BFS) in EDGE and decomposing optimization choices into four orthogonal concerns, we showed that many seemingly disparate graph optimizations arise as composable transformations within a unified framework.
Using GPU BFS as a case study, we systematically traversed a subset of this space. Across 26 categorized variants and over 90 experiments, structured exploration within EEDS yielded up to 1.7× geometric-mean speedups over the best Gunrock baseline on mid- to high-degree-variance graphs.
Our goal is not to claim a single ``best'' BFS implementation, but to demonstrate that graph algorithm design can be organized as principled navigation of a structured transformation space. By making algebraic structure and mapping choices explicit, EEDS exposes optimization dimensions that are typically implicit, enabling systematic reasoning about performance across architectures. We believe this abstraction opens the door to automated exploration for graph algorithms and broader application beyond BFS\@.

\clearpage
\bibliographystyle{./acmart-primary/ACM-Reference-Format}
\bibliography{./bib/all, edge_bfs, temp}
\end{document}